# A Prototype Model of Zero-Trust Architecture Blockchain with EigenTrust-Based Practical Byzantine Fault Tolerance Protocol to Manage Decentralized Clinical Trials

Ashok Kumar Peepliwal[1*], Hari Mohan Pandey[2], Surya Prakash[3], Anand A. Mahajan[4], Sudhinder Singh Chowhan[5], Vinesh Kumar[6], Rahul Sharma[7]

(apeepliwal@gmail.com, profharimohanpandey@gmail.com, suryayadav8383@gmail.com, is anand_mahajan@yahoo.com, sudhinder@iihmr.edu.in, vineshkc@gmail.com, rahulsharma@iihmr.edu.in)

**Abstract:** The COVID-19 pandemic necessitated the emergence of decentralized Clinical Trials (DCTs) due to patient retention, accelerate trials, improve data accessibility, enable virtual care, and facilitate seamless communication through integrated systems. However, integrating systems in DCTs exposes clinical data to potential security threats, making them susceptible to theft at any stage, a high risk of protocol deviations, and monitoring issues. To mitigate these challenges, blockchain technology serves as a secure framework, acting as a decentralized ledger, creating an immutable environment by establishing a zero-trust architecture, where data are deemed untrusted until verified. In combination with Internet of Things (IoT)-enabled wearable devices, blockchain secures the transfer of clinical trial data on private blockchains during DCT automation and operations. This paper proposes a prototype model of the Zero-Trust Architecture Blockchain (z-TAB) to integrate patient-generated clinical trial data during DCT operation management. The EigenTrust-based Practical Byzantine Fault Tolerance (T-PBFT) algorithm has been incorporated as a consensus protocol, leveraging Hyperledger Fabric. Furthermore, the Internet of Things (IoT) has been integrated to streamline data processing among stakeholders within the blockchain platforms. Rigorous evaluation has been done for immutability, privacy and security, mutual consensus, transparency, accountability, tracking and tracing, and temperature–humidity control parameters.

## 1. Introduction

Human subjects are used in clinical trials to test novel medications or complementary therapies to find answers to research problems. However, there are certain problems with how clinical trials are conducted, including delays in receiving regulatory permission, patient selection and retention, data security and privacy, site management, and data manipulation. On the other hand, electronic data capture allows for better control over data fabrication, but recording and reporting data at the global level is time-consuming. Furthermore, in traditional clinical studies, patient retention is difficult [1].

Decentralized Clinical Trials (DCTs) are increasingly embraced to mitigate many possible limitations encountered in traditional clinical trials, such as operational hurdles at sites, difficulties in recruiting and retaining patients, and the need for expedited data access and drug approvals [2,3,4].

In contrast to traditional clinical trials, the management of DCTs effectively tackles the challenge of patient retention by allowing patients to stay in their homes. Real-time data collection via wearable devices minimizes data manipulation while enabling the timely resolution of operational issues that arise during a trial [5].

Hirano et al. [6] underscored the effectiveness of DCTs. They afford the chance to construct highly detailed patient profiles concerning specific treatments and facilitate the analysis of treatment impacts in alignment with clinical trials. Regulators worldwide have endorsed DCTs as vital components of the clinical trial landscape, guiding in integrating remote features and digital endpoints into studies. Patients have voiced satisfaction with the transition to virtual care and communication methods [7].

While DCTs have demonstrated significant enhancements in clinical trials, they encounter several challenges. These include the risk of single-point failure in central data centers, the necessity for trust among stakeholders, scalability issues at the global level, potential compromises of data authenticity, transparency, and confidentiality concerns, the high cost of continuous clinical data recording ($24\times7$), and the subsequent complexities in archiving compared to traditional clinical trials. Notably, advancements in communication technologies such as 5G networks, IoT, blockchain, and zero trust architecture have been instrumental in addressing these challenges associated with DCTs. They have effectively countered these obstacles, fostering the realization of a digital landscape characterized by comprehensive perception and deep interconnectivity, thereby enhancing the conductance of clinical trials.

Refs. [5, 6, 7, 8] reveals the relevance of blockchain technology in clinical trials. Furthermore, EigenTrust-Based Practical Byzantine Fault Tolerance (T-PBFT) improves the operational scalability of overseeing DCTs on a global scale, particularly when dealing with large patient populations in millions. Blockchain operates by distributing blocks or nodes across its decentralized ledger network, where each node receives, processes, and verifies entries while archiving modifications.

Blockchain, Hyperledger-Fabric, and T-PBFT, which are all based on a zero-trust architecture, operate within distributed computer networks and chronologically store data throughout activities. The introduction of a blockchain-based zero-trust model aims to eliminate single-point failures in central data centers and maintain the authenticity, reliability, accuracy, scalability, transparency, and confidentiality of stored clinical data [2]. Integration of blockchain with the Internet of Things (IoT) enables researchers to conduct DCTs realistically while adhering to study protocol procedures, ensuring patient safety, compliance with ICH-GCP standards, and other relevant regulatory guidelines. The Internet of Things, or "things", is a network of physical objects with sensors, software, and other technologies implanted in them that allow them to communicate and share data with other systems and devices over the internet [16].

Wearable sensors that give real-time health data from trial participants are one-way IoT devices that can help collect distant data. To improve the effectiveness of DCT activities, the IoT can also facilitate interoperability, machine-to-machine connectivity, information exchange, and data transfer [18].

These technologies facilitate the remote execution of DCT-related tasks on an individual patient basis, record DCT-generated data [5] with timestamps, expedite the accessibility of patient Case Record Forms (CRFs), promptly resolve Data Clarification Forms (DCFs), accelerate the research process, expedite regulatory approvals, and ensure data reliability throughout the trial.

A thorough analysis of T-PBFT and a comparison with alternative Byzantine fault-tolerant consensus algorithms revealed that T-PBFT enhances scalability and fault tolerance, reduces the occurrence of view shifts, and simplifies communication complexity, as per theoretical investigations. This research proposed a pioneering model for integrating blockchain, IoT, Hyperledger Fabric, and T-PBFT to facilitate the seamless operation of DCTs worldwide.

The aforementioned discussion leads to the investigation of certain research questions (RQs), which are highlighted below:

RQ1  Is it possible to integrate DCTs at the global level?
RQ2  How will blockchain, IoT, and Hyperledger Fabric systems work on zero-trust architecture?
RQ3.  How does T-PBFT enhance the scalability of DCT?

The key contributions of this paper are highlighted as follows:

1. First, we present the integration of DCTs with blockchain.
2. Second, we discuss the functions of blockchain, the IoT (as wearable devices), and the Hyperledger fabric with a zero-trust architecture system.
3. Third, enhancing T-PBFT scalability for DCTs.

Typically, we present a structured flow outlining the prototype of the Zero-Trust Architecture Blockchain (z-TAB) model. It encompasses a literature review and a reasoned approach to model development, utilizing blockchain and Hyperledger Fabric systems to manage the private blockchain. It incorporates IoT devices for remote access to DCT data via wearable devices, smart functions for automated process execution, and T-PBFT to ensure mutual consensus among operational nodes. The applicability of the z-TAB model to DCTs is discussed, along with the evaluation of the developed model based on specific operational parameters of DCTs [4].

The rest of the paper is organized as follows: Section 2 presents related work. The rationale of z-TAB model development is described in Section 3. Section 4 explains the blockchain-based zero-trust architecture with Hyperledger and T-PBFT, and Section 5 discusses the z-TAB model. The applicability of the z-TAB model for DCTs is presented in Section 6. Section 7 highlights the z-TAB model evaluation in the operation management of DCTs; and concluding remarks, implications, and directions for further research are presented in Section 8.

## 2. Related work

Gergova et al. [9] evaluated the integration of decentralized components into clinical trials across Europe, highlighting the need for meticulous, customized consideration. European nations increasingly favor a hybrid clinical trial model, blending onsite visits with decentralized elements, and viewing it as superior to the traditional model. However, the application of national regulations often lacks specificity for such scenarios. Jakkula et al. [10] stressed DCTs' operational feasibility and benefits, citing higher participation rates, improved compliance, reduced dropout rates, and faster completion times. DCTs align with the industry's pursuit of low-risk, high-yield trials, offering the convenience of home participation and continuous operation with real-time data and patient-centric focus.

De Brouwer et al. [11] proposed employing edge computing, a zero-trust architecture, and federated computing in DCTs, alongside supportive policies and regulations, to ensure user safety and accelerate clinical research. de Jong et al. [12] identified regulatory barriers and benefits of implementing DCTs within the European Union, highlighting concerns regarding investigator supervision and participant safety in restricted physical interaction scenarios.

Kouicem et al. [13] examined security and privacy solutions for the IoT, emphasizing the potential of blockchain and software defined networking to enhance flexibility and scalability.

Omar et al. [14] discussed blockchain-based solutions in clinical trials, addressing challenges in integration. Krishnamurthi et al. [15] explored consensus algorithms and challenges in blockchain technology. Li et al. [16] studied blockchain for securing transportation processes. Sandner et al. [17] integrated blockchain, IoT, and AI, focusing on data collection, infrastructure, and security. Hosen et al. [18] proposed a transaction validation protocol for secure IoT networks using blockchain and software defined networking. de-Melo-Diogo et al. [19] illustrated blockchain's role in overseeing clinical trials. Feng et al. [20] introduced a blockchain-based identity storage system for secure data updates. Gussinklo et al. [21] developed BlockTrial, a blockchain-powered clinical trials management system. Izmailova et al. [22] assessed wearable devices in drug development trials. Awan et al. [23] researched a secure IoT architecture utilizing blockchain. Wang et al. [24] integrated zero-trust security into medical systems. Liu et al. [25] optimized consensus processes in group communication. Gao et al. [26] introduced T-PBFT, a practical Byzantine fault tolerance consensus method utilizing the EigenTrust model. A summary of related research is presented in Table 1.

**Table 1**

Summary of related research work

| Reference | Research objectives | Proposed research |
|---|---|---|
| [9] | This study examines European nations' experiences and methods for implementing decentralized components and a hybrid strategy for conducting clinical trial procedures and activities. | Using email correspondence, a questionnaire poll was sent to all European countries between December 2020 and February 2021, and the data were analyzed. |
| [10] | To conduct a review on clinical trials transformation initiative-decentralized clinical trials: | Clinical trial sponsors can now employ best practices and workable solutions to these problems disclosed by the Clinical Trial Transformation Initiative (CTTI). |
| [11] | To explain how technologies like federated computing, edge computing, and zero-trust environments affect Decentralized Clinical Trials (DCTs). | Digital Health Technologies (e.g., smart devices, new wearables, and environmental sensors) facilitate multiple trial-related activities: Stakeholder communication, patient enrolment, recruitment, informed consent, and continuous data access. |
| [12] | To determine the prospects and regulatory obstacles for DCT deployment in the European Union. | The research was conducted in semi-structured interviews with twenty European regulators. Respondents suggested hybrid clinical trials that combine decentralized and onsite components. |
| [13] | To combine the digital and physical realms seamlessly into a unified ecosystem to create a new intelligent internet era. | A thorough top–down analysis of the most recent IoT security and privacy proposals of emerging methods like blockchain and software-defined networking can improve the flexibility and scalability of IoT security and privacy. |
| [14] | To address strict data management problems in Clinical Trials (CTs) (such as patient recruitment, ongoing monitoring, data management, data analytics, and accurate reporting). | This survey observations are on the blockchain's acceptance in CTs. It shared information on ongoing efforts to implement blockchain technology in CTs. |

| [15] | To identify different consensus algorithms, blockchain challenges, and their scope. | The study examined the fundamental idea behind blockchain technology and a few mining methods, consensus issues, consensus algorithms, and performance-based comparison algorithms. |
|---|---|---|
| [16] | To prevent privacy leakage throughout the entire transportation process from sender to receiver. | Eleven techniques for processing IoT data with blockchain technology were compiled to guard against privacy breaches during the full sender-to-receiver procedure. |
| [17] | To convergence of blockchain, IoT and AI. | Blockchain technology, in conjunction with IoT and AI, will lead to a new era of digitization. |
| [18] | To suggest a secure distributed Internet of Things network's transaction validation methodology using blockchain technology. | Proposed a transaction validation protocol for secure IoT networks using blockchain and software defined networking. |
| [19] | To map the current utilization of blockchain systems in clinical trials. | By providing precise, certified data, blockchain ensures data security in situations where the data processing process is more transparent and results in tamper-proof clinical trials that are more credible and dependable. |
| [20] | Enhancing the system's security, efficiency, and stability can guarantee railway transportation's safety and reliability. | Introduced blockchain-enabled zero trust-based authentication scheme and Merkle tree to develop a distributed identity storage system that ensures rapid, discreet, and trustworthy data updates while enhancing the effectiveness of authentication. |
| [21] | To develop a proof-of-concept system and investigate how blockchain technology can assist in managing clinical trial data. | Described BlockTrial, a system that uses a Web-based interface to allow users to run trials-related smart contracts on an Ethereum network. |
| [22] | To facilitate further evaluation and adoption of wearable devices in clinical trials. | The study emphasized the logistical and methodological factors that should be considered when conducting clinical trials, along with the essential components of clinical and analytical validation within the particular context of use (COU). |
| [23] | To monitor and enable device-to-device communications with varying degrees of access-controlled mechanisms in response to environmental factors and device behavior. | Research has covered the main threats and weaknesses posed by cyber threats in smart environments using a novel secure framework called ZAIB (zero-trust and ABAC for IoT using blockchain). |
| [24] | To ensure the security of medical information systems. | The study integrated the medical system with the zero-trust security system to present a zero-trust medical security system. Furthermore, to enhance the security of medical equipment and data, under zero-trust conditions (ABEAC). This model was developed using the role-based access control (RBAC) model, user behavior risk value, and trust calculations. |
| [25] | To improve practical Byzantine fault tolerance (Practical Byzantine Fault Tolerant consensus algorithm based on reputation, RPBFT) for the problems of high communication complexity, poor scalability, and random selection of master nodes of consensus algorithm of the consortium chain. | A simulation and performance testing system based on practical Byzantine fault tolerance (Practical Byzantine Fault Tolerant consensus algorithm based on reputation, RPBFT) is built to prove the scheme's effectiveness and usability through simulation experiments. |

| [26] | To analyze T-PBFT and compare it with the other Byzantine fault-tolerant consensus algorithms. | A novel optimized practical Byzantine fault tolerance consensus algorithm based on the EigenTrust model, T-PBFT, is a multi-stage consensus algorithm. |

## 3. Rationale of z-TAB model development

Unlike traditional clinical trial methods [27,28], DCTs provide heightened global security and transparency throughout trial execution. They enable remote patient access and real-time retrieval of clinical data while upholding the principles of attributable, legible, contemporaneous, original, accurate, and complete (ALCOA) documentation [29]. In DCTs, patients stay connected through wearable devices or patient engagement tools, allowing them to relocate without straying from trial protocols. Patient data are seamlessly captured via these wearable devices, ensuring alignment with electronic data capture (EDC) systems or clinical data management systems (CDMSs) before transmission to the trial sponsor, as illustrated in Fig. 1.

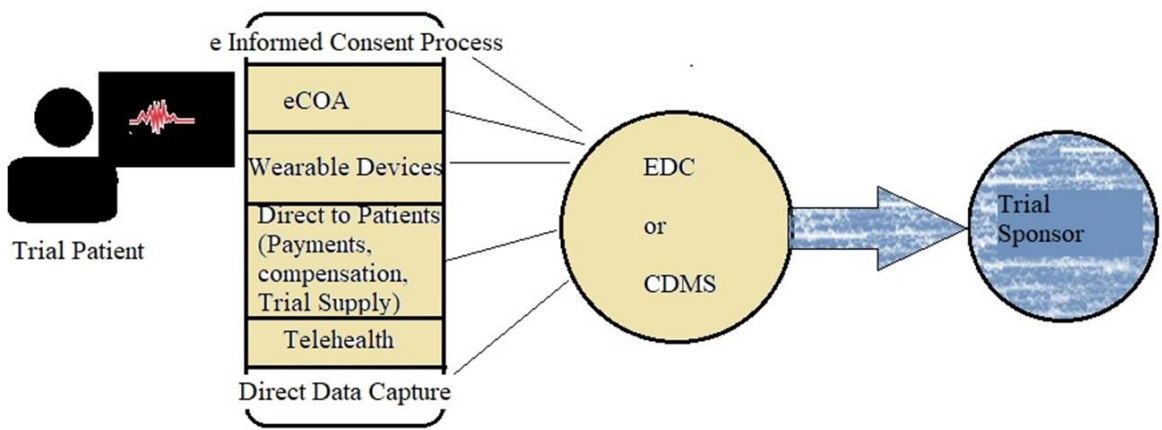

**Fig. 1.** Decentralized clinical trials [30].

Accessing decentralized trial data occurs at specified intervals and can be continuously monitored in a controlled fashion [31]. Due to their larger datasets than conventional trials, decentralized trials may accommodate broader variability tolerance, potentially leading to a higher likelihood of missing data [32]. The emerging integration of IoT and blockchain technologies holds the potential for establishing zero-trust architecture in DCTs, ensuring the integration and security of trial-generated data. These trials rely less on intermediaries and specialized research facilities for data collection.

Integrating clinical data from IoT devices, reflecting real-world scenarios, can provide additional context for online and in person clinical encounters. IoTs, encompassing applications and medical equipment communicating over Internet networks, facilitate access to healthcare IT systems. Wi-Fi-enabled medical devices allow for machine-to-machine communication. Coupled with technologies like blockchain, these approaches aim to improve patient comfort, compliance, and the speed of real-time data collection compared to traditional clinical trial methods [33].

The patient assessment activities outlined in Table 2 are primarily conducted virtually, except for in-person tasks, utilizing various IoT devices. Data from patients are directly captured through wearable devices, either in the form of data signatures or hash values within the z-TAB model.

**Table 2**

Assessment activities of patients (virtual and in person mode).

| Patient study visit No. | Assessment parameter | Mode | | Coordinating point for activity |
|---|---|---|---|---|
| | | Virtual | In Person | |
| 1 | Patient screening/Identification | Virtual | No | PI and coordinator |
| 2 | Informed consent process | Virtual | No | PI and coordinator |
| 3 | Pre-study assessment (Physical examination, Pregnancy test, Vitals, ECG, Laboratory assessment*) | Virtual | *In person | Phlebotomist and laboratory personnel |
| 4 | Physical examination, Vitals (Temp/BP), ECG | Virtual | No | PI and coordinator |
| 5 | Physical examination, Vitals (Temp/BP), ECG | Virtual | No | PI and coordinator |
| 6 | Physical examination, Vitals (Temp/BP), ECG | Virtual | No | PI and coordinator |
| 7 | Physical examination, Vitals (Temp/BP), ECG | Virtual | No | PI and coordinator |
| 8 | Laboratory assessment* | No virtual | *In person | Phlebotomist and laboratory personnel |
| 9 | Physical examination, Vitals (Temp/BP), ECG | Virtual | No | PI and coordinator |
| 10 | Physical examination, Vitals (Temp/BP), ECG | Virtual | No | PI and coordinator |
| 11 | Physical examination, Vitals (Temp/BP), ECG | Virtual | No | PI and coordinator |
| 12 | End of study [(Physical Examination, Vitals (Temp/BP), ECG, Laboratory assessment*] | Virtual | *In person | PI and coordinator, Phlebotomist and laboratory personnel |

* Activity that could be completed in person, not virtually.

In this context, a zero-trust architecture is employed, functioning within both external and internal network environments, and verifying transactions before broadcast each time [32]. Consequently, a model for operating DCTs on a global scale utilizing z-TAB is under development [23,33]. z-TAB, in conjunction with Hyperledger Fabric and T-PBFT as the consensus protocol, is applied to facilitate data transfer from patients to principal investigators and other DCT stakeholders, ensuring data integrity and security within this framework. The model undergoes evaluation on criteria including data immutability, mutual consensus, transparency, accountability, temperature and humidity control within the supply chain, IMP traceability, privacy, and security, with the aim of enhancing its authenticity and acceptability [36].

## 4. Blockchain-based Zero-Trust Architecture with Hyperledger Fabric and T-PBFT

*4.1. Blockchain and Hyperledger Fabric architecture in DCTs*

The inherent immutability of blockchain technology can enhance the security of the zero-trust model, potentially enabling blockchain to identify, validate, and grant access to trusted models [37]. Blockchain-enabled zero-trust security can isolate connections, detect suspicious online transactions, and restrict user access [37]. Blockchain operates as a decentralized ledger technology, where blocks are sequentially added in a chronological manner. In DCTs, data can be accessed within a blockchain framework. Blocks representing DCT stakeholders are interconnected in a timestamped manner, forming a decentralized and tamper-proof chain of data. This cryptographically secured data source holds promise for addressing key challenges in healthcare, particularly in multicentric clinical trials, where data integrity, traceability, and transparency are paramount [38].

Clinical data are collected, stored, and transferred during DCTs using IoT devices. These data can be stored on a blockchain platform, facilitating interconnected sharing among patients, principal investigators, regulators, contract research organizations (CROs), and sponsors [39, 40]. This study used Hyperledger Fabric to construct a decentralized system for operational management within the z-TAB paradigm. Hyperledger Fabric's design supports fully decentralized blockchain networks, with the private blockchain framework developed by the Linux Foundation [Fig. 2] [41].

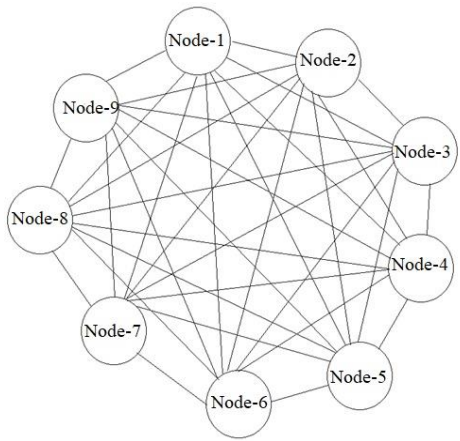

**Fig. 2.** Hyperledger fabric system.

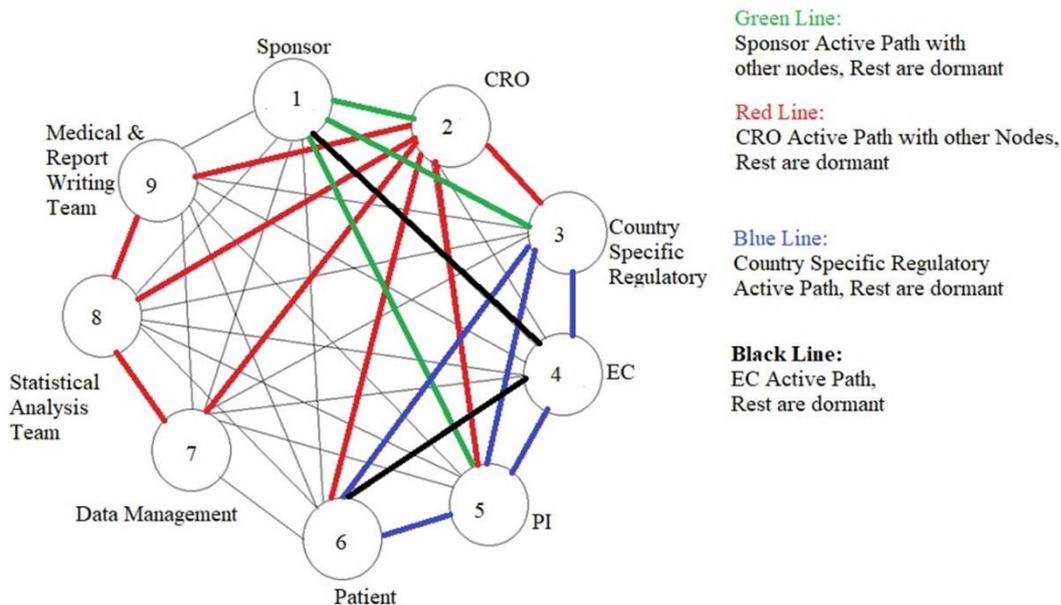

**Fig. 3.** Private channels on the Hyperledger fabric pluggable architecture.

The system architecture is highly adaptable, allowing for the integration of additional functionalities such as membership services, identity management, encryption, and consensus protocols. Within the private network, a variety of nodes are present, including those representing Contract Research Organizations (CROs), countries, ethics committees, principal investigators, patients, data management entities, statistical analysis units, medical teams, and report-writing entities. Furthermore, the network encompasses a smart contract (chaincode), a ledger containing a state database, and a transaction log.

In a Hyperledger Fabric system, there are different types of nodes: client nodes (representing patients), which initiate data transactions; peer nodes (associated with private channels), which are responsible for maintaining the ledger of transaction data; and ordered nodes, facilitating communication and transaction order maintenance [41]. Fig. 3 illustrates nodes 1 through 9 within the z-TAB framework, where clinical data transactions occur during DCTs. Peer nodes within private channels continuously update the ledger upon receiving data directly from patients. Various private channels operate within the fabric, as outlined in Table 3.

**Table 3**
Nodes of private channels.

| Private Channel Number | Nodes of private channel | Name of nodes on specific private channel |
|---|---|---|
| 1 | Node-**1**,2,3,4,5 (Green Bold Line) | **Sponsor**, CRO, Country Specific Regulatory, EC, PI |
| 2 | Node-**2**,3,5,6,7,8,9(Red Bold Line) | **CRO**, Country Specific Regulatory, PI, Patients, Data Management, Statistical Analysis Team, Medical & Report Writing Team |
| 3 | Node-**4**,1,5,6 (Black Bold Line) | **EC**, Sponsor, PI, Patients |
| 4 | Node-**5**,1,3,6 (Blue Bold Line) | **PI**, Sponsor, Country Specific Regulatory, Patients |

*Note:* Node, which is in bold character, is the Client node, which initiates the transaction in a particular channel.

In Fig. 4, the client node (associated with a private channel) submits a transaction proposal to the orderer node, which sends the data transactions to the endorsers. Another peer node within the channel maintains the ledger of clinical data transactions and commits the transaction. Upon receiving the ordered state from the orderer, the peer node updates the ledger. The peer node acts as an endorser before a transaction is submitted to the orderer. The orderer node verifies the endorsement before delivering the data transaction to the peer nodes.

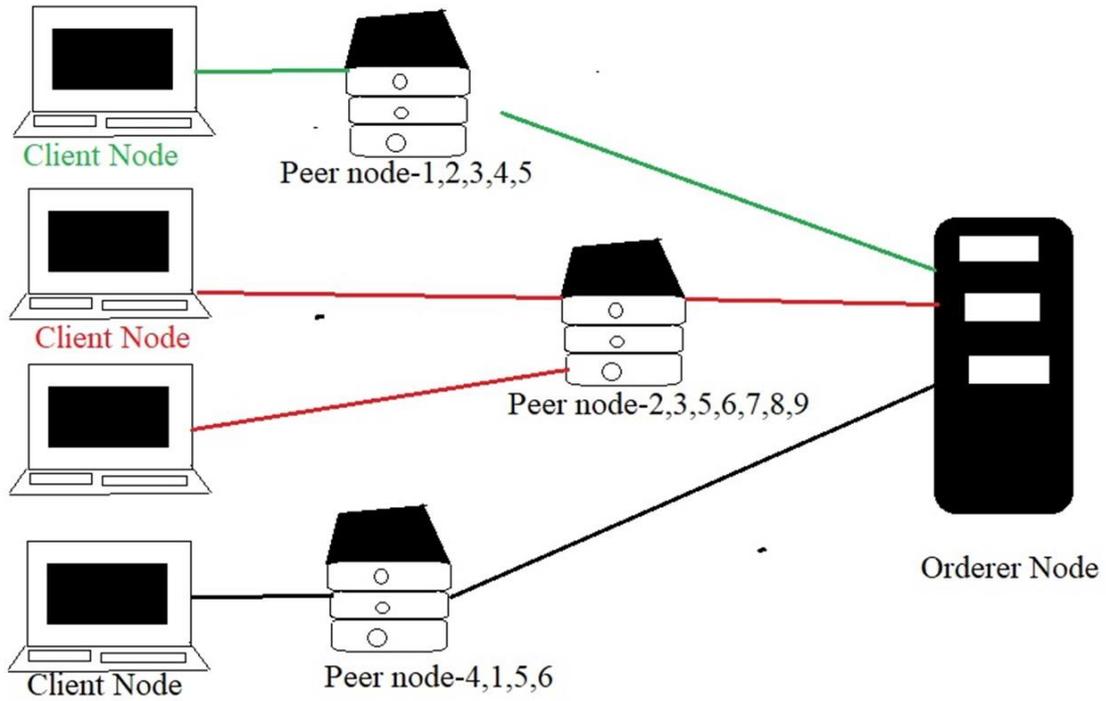

**Fig. 4.** Hyperledger fabric system architecture [42].

The private channel network consists of peer nodes, which also function as client and endorser nodes, as well as an orderer node. Nodes 1, 2, 3, 4, and 5 belong to private channel No. 1, sharing a data ledger and operating on the same smart contract. In contrast, channels No. 2 (nodes 2, 3, 5, 6, 7, 8, 9) and No. 3 (nodes 4, 1, 5, 6) have distinct ledgers and work on separate smart contracts. The orderer node's role involves proposing transactions, validating endorsements, IDs, signatures, and broadcasting transaction messages to peer nodes.

Transactions on the blockchain are governed by smart contracts, referred to as chaincodes in the Hyperledger Fabric system. Rules are encoded as functions within chaincode, and Hyperledger Fabric enforces endorsement policies where transactions are verified by predetermined endorsing nodes within a private channel after being initiated by a client [43]. The orderer node ensures the validity of messages from each endorser by confirming sufficient valid endorsed signatures and simulating data transactions.

Once collected, data transactions are distributed to other peers within private channels as a new block. Participants within the private network are enrolled by a trusted Membership Service Provider (MSP), which assigns digital identities to all blockchain nodes on the network, whether they serve as peers, orderers, or clients.

### 4.2. Blockchain and IoT-based modeling for DCTs

DCTs leveraging blockchain and IoT infrastructure aim to overcome the challenges faced by conventional data management systems in multi-site clinical trials. We designed our DCTs using Hyperledger Fabric, utilizing built-in capabilities such as private networks, private channels, and smart contracts. Specific network routes are activated during data transfer, while others remain inactive.

This section presents the setup of the Hyperledger Fabric network, the installation of private networks, and the creation of customized smart contracts for each network. Security and privacy are paramount concerns when sharing data over the IoT. Adopting a peer-to-peer architecture is advised, with blockchain technology ensuring privacy in IoT networks. Blockchain controls all activities on IoT data, aiding in detecting and addressing data exploitation.

Blockchain and the IoT revolutionize DCTs, with IoT devices securely storing patient-centric remote data on blockchain-based distributed ledgers via cloud computing [44]. In blockchain, each stakeholder is represented as a node interconnected within the network (Fig. 5). The ledger contains verified transaction proofs, forming an immutable chain. Each node contains various blocks comprising hashes, a list of valid transactions, and the previous block's hash, ensuring the tamper-proof nature of blockchain.

Blockchain is categorized based on the ledger generated during information transactions between peers: public ledger (permissionless framework) and private ledger (permissioned framework) [45].

Blocks serve as digital containers that permanently house data pertaining to network transactions. Each block records any or all the most recent data transactions that have not yet been included in earlier blocks. When a block is "completed," the blockchain proceeds to the next block. Thus, a block acts as a repository for records that, once written, remain immutable and cannot be altered or deleted.

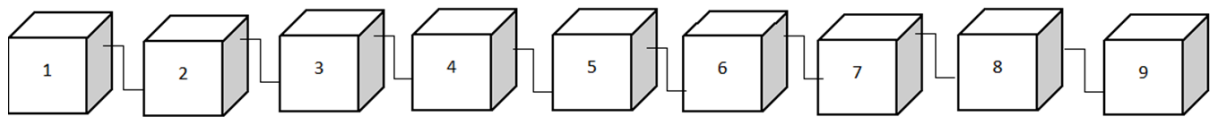

**Fig. 5.** Coupled nodes (Node-1 to Node-9) on the blockchain [46].

This paper adopts a private ledger-based blockchain to ensure and maintain data privacy among stakeholders exclusively. Only verified and preapproved participants are allowed to join a private or permissioned network blockchain, access the ledger, carry out transactions, and take part in consensus techniques like Practical Byzantine Fault Tolerance (PBFT) and Proof of Elapsed Time (PoET) [47].

Node-1, representing the sponsor, updates its ledger with transactional information during DCTs through the smart contract on its private channel. This node serves as the genesis node, storing transactions related to the planning of multicentric trials in its blocks. These blocks are generated on Hyperledger Fabric, a permissioned and open network comprising various nodes that interact to fulfill their designated roles. Fig. 6 illustrates the flow of information transactions among the *n*-nodes (Node-1 to Node-9) within private channels, with other private channels remaining obscured on the blockchain and the data being partitioned.

Activity-based private channels among the nodes enable specific data points to be accessible only to nodes requiring the relevant transactional information. Different clinical trial activities, such as the informed consent process, patient recruitment, trial monitoring, data analysis, and report writing, have their respective private channels on the Hyperledger Fabric-based blockchain, each with a unique method of data transaction [41].

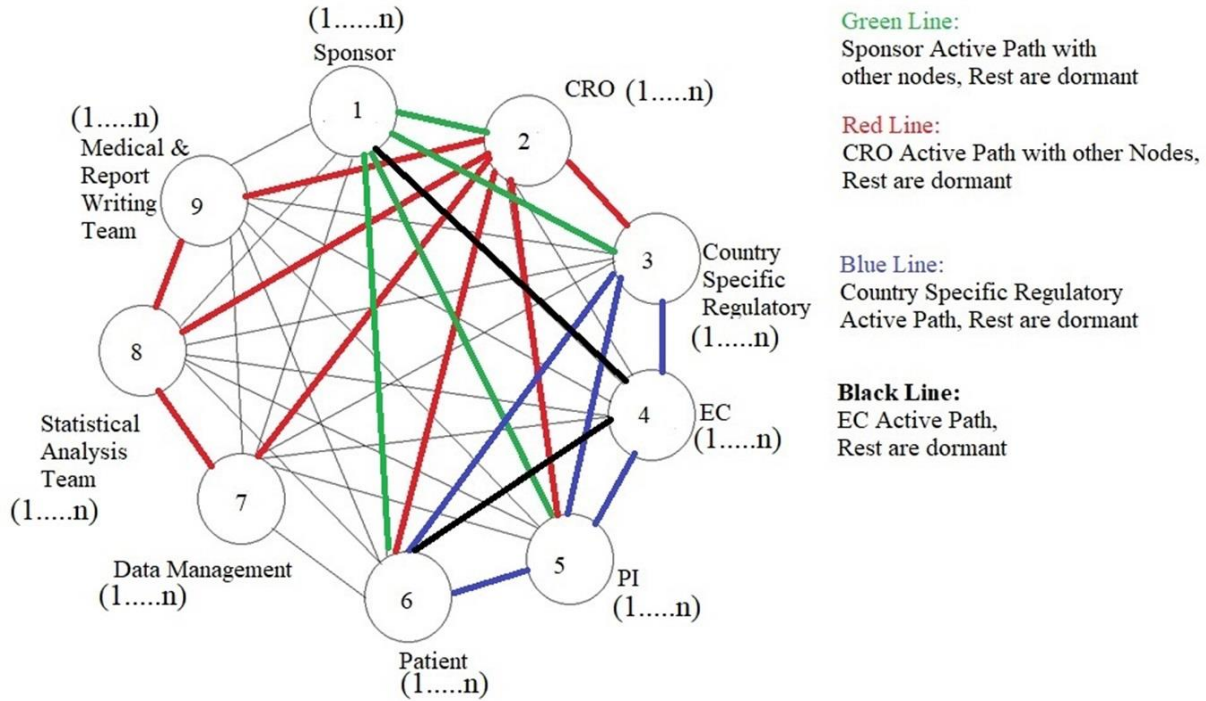

**Fig. 6.** Private channels (Node-1 to Node-9) on Hyperledger Fabric [48].

Numerous sponsors, CROs, regulatory bodies, and other stakeholders could exist on the Hyperledger Fabric. To represent this diversity, they are expressed as numbered entities, ranging from 1 to *n*. For instance, there could be CRO1, CRO2, CRO3, Country regulatory1, Country regulatory2, Country regulatory3, and so forth. This numbering system allows for the definition of active and inactive nodes across different channels (refer to Table 4).

**Table 4**
Private channels active nodes and functions.

| Name of channel | Nodes of the private channel | Active and inactive nodes on a private channel | Functions of channel |
|---|---|---|---|
| Patient enrollment channel | Node-4,5,6 | Active: 4,5,6 Inactive:1,2,3,7,8,9 | Patient identification, screening, recruitment, patient data access through wearable devices |
| Trial monitoring channel | Node-1,5,6 | Active: 1,5,6 Inactive:2,3,4,7,8,9 | Patient status, withdrawals, completion of study, report preparations |
| Clinical data analysis channel | Node-2,7,8 | Active: 2,7,8 Inactive:1,3,4,5,6,9 | Clinical data access, data cleaning, data analysis, and outcome assessment |
| Medical & report writing channel | Node-2,7,8,9 | Active: 2,7,8,9 Inactive:1,3,4,5,6 | Medical and report preparation in desired format |

A multi-site clinical study uses a blockchain-based system with private channels for data management, where each participant maintains a ledger of transactions and uses a member channel's smart contract. It inhibits unauthorized data access and preserves information confidentiality by limiting data transactions to channel members exclusively. The transaction of data is in the form of hash values, which are generated against the text data received through the wearable devices of remotely randomized patients, and the blocks are connected to each other through the hash values (Fingerprint) of clinical trial data.

*4.3. Smart contract function of the blockchain model*

The essence of smart contracts is rooted in blockchain technology. To ensure adherence to the regulations governing clinical trial protocols, smart contracts have become indispensable. These contracts, essentially computer programs or protocols, operate autonomously, executing tasks such as self-execution, self-administration, self-validation, and self-impediment when specific conditions are fulfilled within a blockchain environment, all without delays. Powered by Distributed Ledger Technology (DLT), smart contracts automate processes and facilitate global data storage across servers, with stored information as the bedrock for transaction verification [49].

A smart contract comprises essential elements such as value, address, function, and state. Upon receiving a transaction as input, the relevant code is executed, triggering an output event and subsequent changes in state based on functional logic. In DCTs, where multiple stakeholders engage in data transactions, smart contracts play a crucial role in ensuring that data flow through the legitimate pathway of the Hyperledger Fabric system. These programs can be customized to encompass a range of functions tailored specifically for conducting clinical trials. The activation of smart contract features is facilitated through interaction with an application interface by blockchain users [50].

The matching function ensures that each data transaction request originates from an authorized user for an approved channel, data type, and timeframe, thereby enabling precise access control. Before deployment, stakeholders collectively establish the terms of the smart contract, outlining triggering circumstances for contract execution, protocols for state transitions (in compliance with DCT requirements such as ICH-GCPs, ECs, protocols, and other relevant regulatory standards), and mechanisms for holding parties accountable for contract breaches. The smart contract is subsequently encoded as code and published onto the blockchain. Once the predetermined conditions are met, the smart contract activates and executes automatically.

The Sponsor, identified as Node-1 within the blockchain, represents a pharmaceutical organization funding DCTs across various countries and overseeing clinical trial operations. The Sponsor delegates significant responsibilities to Contract Research Organizations (CROs), which collaborate with different principal investigators to conduct clinical studies in hospitals or research centers. Node-1, acting as the Sponsor, assumes the duty of implementing essential clinical trial prerequisites by the International Council for Harmonisation of Technical Requirements for Pharmaceuticals for Human Use - Good Clinical Practice (ICH-GCP) guidelines. These prerequisites encompass protocol development, patient indemnity, Informed Consent Forms (ICFs), investigator brochures, monitoring teams, safety and risk control plans, statistical plans, data access, and monitoring plans. Smart contract functions are programmed to execute automatically within the blockchain model once the specified conditions are satisfied, subsequently updating the ledger on the blockchain and replicating the data onto other authorized blocks (nodes 2–9). The procedural steps of the smart contract process on z-TAB are delineated.

**Step 1**. Once the Sponsor drafts a contract outlining prerequisite conditions in code format, it is transmitted to subsequent stakeholders to fulfil DCT functions throughout the blockchain system. Upon completion of the agreement and dissemination of information, other blocks validate receipt of the distributed ledger.

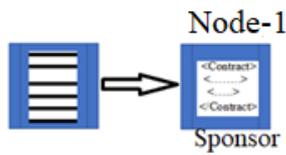

**Fig. 7.** Sponsor transfers contract in the form of codes

**Step 2**. The code is replicated from Node-2 to Node-9 and saved across the blockchain stakeholders.

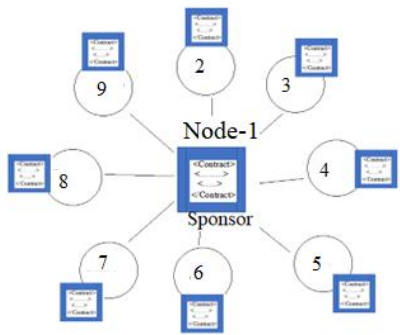

**Fig. 8.** Code replication on DCT stakeholder's nodes

**Step 3**. Every computer linked to the blockchain network executes the code and implements it. When a condition defined for DCTs is satisfied and verified by each block on the blockchain network, the associated transaction is executed.

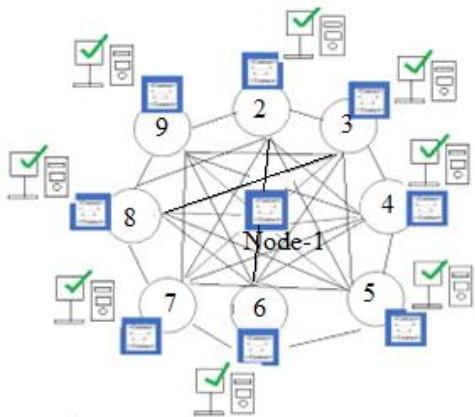

**Fig. 9.** Computers in the network check the correctness of DCT conditions, satisfied and validate the data transaction.

The smart contract among the nodes allows the DCT activities to be performed in sequential ways on a pre-determined specific condition met as per the protocol, ICH-GCP, and other applicable regulatory requirements. Node specific trial activities are controlled by the smart contracts on an automatic route from node-1 to node-9.

| Node-1 (Genesis Node) | Node-1 Hash | Node-1 (Sponsor) |
|---|---|---|
| **Header** | 0000000000176e516649db9d2eb4bfb42ed34f7e32626ac3837182c58ff79f7e | A blockchain is a distributed database that maintains an ever-growing list of ordered records, known as blocks, interconnected through cryptography. Node-1, also known as the sponsor, initiates the blockchain with a genesis block, to which subsequent blocks are sequentially added. The blocks are linked using cryptographic hash functions, essential for data verification and utilized within smart contracts. Each block contains transaction details such as Node-1's (Clinical Study related activities), a timestamp, Merkel Root, Nonce, and a cryptographic hash of the preceding block. The genesis block differs from others by having two additional leading hex zeros in its hash (0000000000176e516649db9d2eb4bfb42ed34f7e32626ac3837182c58ff79f7e), making it unique. This hash encapsulates the clinical trial activity data of Node-1, accessible to the sponsor and other authorized entities. |
| Previous Hash (000000000) | | |
| Time Stamp (DDMMYYYY, 00:00:00 GMT) | | |
| Nonce | | |
| Merkle Root (Node-1 Transacted data) | | |
| **Transaction Data (Clinical Study related activities)** | | |
| • Clinical Study Protocol<br>• Investigator Brochure<br>• Informed Consent Document<br>• IMP Management<br>• Trial Management and Monitoring<br>• Financing<br>• Safety Monitoring and Reporting<br>• Clinical Study Report Preparation<br>• FDA Report Submission and Approvals | | |

Clinical trial activities from Node-2 to Node-9 adhere to the approved protocol. As illustrated in Fig. 10, each node updates information within a sequence of blocks, accomplishing protocol-specific tasks virtually. However, physical collection of biological samples (such as blood, urine, saliva, etc.) is required for investigations.

| Node-2 | Node-2 Hash | Node-3 | Node-3 Hash | Node-4 | Node-4 Hash | Node-5 | Node-5 Hash |
|---|---|---|---|---|---|---|---|
| **Header**<br>Previous Hash (Node-1 Hash)<br>Time Stamp (DDMMYYYY, 00:00:00 GMT)<br>Nonce<br>Merkle Root (Node-2 Transacted data)<br>**Transaction Data (CRO related activities)**<br><br>•Agreement with Sponsor<br>•Site and PI Selection<br>•Logistics and Material shipment<br>•Site Initiation, Training to PI, Monitoring, and close-out<br>•Data access and control<br>•Protocol, ICH-GCP and Regulatory Compliance<br>•SAE reporting | 258a60f8ca546a02e7cbe77bf7bbd2ee19f0947b592c161e75d98a39157a35c3 | **Header**<br>Previous Hash (Node-2 Hash)<br>Time Stamp (DDMMYYYY, 00:00:00)<br>Nonce<br>Merkle Root (Node-3 Transacted data)<br>**Transaction Data (Country regulatory related activities)**<br><br>•Study Protocol Application<br>•Application Review<br>•Protocol related queries<br>•Query resolvement<br>•Protocol approvals<br>•Pharmacovigilance<br>•Patient Safety<br>•Country Specific regulatory compliance<br>•Protocol Amendments | 8ba54502b2a84d27b1e7e64ba338ac0ecbe906652ca06b2bf6cb503ab71d7e45 | **Header**<br>Previous Hash (Node-3 Hash)<br>Time Stamp (DDMMYYYY, 00:00:00 GMT)<br>Nonce<br>Merkle Root (Node-4 Transacted data)<br>**Transaction Data (EC related activities)**<br><br>•EC Composition and SOP<br>•EC meeting call<br>•Protocol Review<br>•Study Approvals<br>•SAE report and compensation<br>•Regulatory compliance<br>•EC Data Archival | 656a4605114d82d6275d2d585d3d8f3cd993c22d74cfd589fea8815f7cc0b0515 | **Header**<br>Previous Hash (Node-4 Hash)<br>Time Stamp (DDMMYYYY, 00:00:00 GMT)<br>Nonce<br>Merkle Root (Node-5 Transacted data)<br>**Transaction Data (PI related activities)**<br><br>•Study Protocol submission to EC<br>•Communication to EC, sponsor and regulatory<br>•Patient Informed Consent<br>•Patient screening, recruitment, and enrolment<br>•Financing<br>•Safety Monitoring and Reporting<br>•Clinical Study Report Preparation<br>•FDA Report Submission and Approvals | bcec83ec83bc36dc17703f3b4d694534 3e2895f0118b9e67adfdac34 27fd612a0 |

| Node-6 | Node-6 Hash | Node-7 | Node-7 Hash | Node-8 | Node-8 Hash | Node-9 | Node-9 Hash |
|---|---|---|---|---|---|---|---|
| **Header**<br>Previous Hash (Node-5 Hash)<br>Time Stamp (DDMMYYYY, 00:00:00 GMT)<br>Nonce<br>Merkle Root (Node-6 Transacted data)<br>**Transaction Data (Country Site Patient related activities)**<br><br>•Patient inclusion/exclusion compliance<br>•Biological sample (blood, urine, saliva etc.) collection<br>•Patient Randomization<br>•IMP administrations<br>•Wearable Devices Use<br>•Mobile Application and Internet<br>•24 x7 data access and monitoring<br>•SAE reporting | 08f24d8fc1ff318b24ff246 2e76280219d4dea0708df3 c6c9d7f02fb82c76a06 | **Header**<br>Previous Hash (Node-6 Hash)<br>Time Stamp (DDMMYYYY, 00:00:00 GMT)<br>Nonce<br>Merkle Root (Node-7 Transacted data)<br>**Transaction Data (Data Management related activities)**<br><br>•CRF designs, Annotation and Data collections<br>•Data Review and DCF generation<br>•DCF query resolution<br>•Data Entry<br>• Training to Data Management Team<br>•Data Validation<br>•Medical Coding<br>•Database Lock | 39a668efe112b64284297bc455709712c4c68a45e045329ec46ff9e318da9fb3b | **Header**<br>Previous Hash (Node-7 Hash)<br>Time Stamp (DDMMYYYY, 00:00:00 GMT)<br>Nonce<br>Merkle Root (Node-8 Transacted data)<br>**Transaction Data (Stattistical Analysis related activities)**<br><br>•Data Access for analysis<br>•Data Analysis<br>•Statistical methods and study conclusion<br>•Missing data<br>•Data Sharing to Medical Writers | 0d53b2407ce67d06d2817a f122851ed92da6593f57f97 a56e2d25d59451e0aa71 | **Header**<br>Previous Hash (Node-8 Hash)<br>Time Stamp (DDMMYYYY, 00:00:00 GMT)<br>Nonce<br>Merkle Root (Node-9 Transacted data)<br>**Transaction Data (Medical and Report Writing)**<br><br>• Medical Writing<br>• Report format<br>• Protocol procedures<br>•Data Access<br>•Report Writing | ac62cbf3228708 6dd79f3f5a9c892b2514aef4bbcccfa48a61b3b085afb4f00 |

**Fig. 10.** Clinical study protocol-specific activities on the blockchain

Patients undergo electronic screening from an existing database, and prospective participants are recruited based on specific conditions outlined in the inclusion and exclusion criteria (such as gender, age, pre-existing conditions, and medical history) according to the clinical trial protocol. Upon patient eligibility determination, the Principal Investigator (PI) virtually obtains consent, with the data stored in Node-6. The enrollment smart functions validate each study criteria condition before the data are appended to the DCT Hyperledger Fabric on the blockchain. Other authorized stakeholders of the DCT have read or write access to this ledger but cannot make changes. Smart contract functions facilitate cryptographic communication among stakeholders, utilizing hash functions to generate hash values for input transaction data.

*4.4. Merkle tree of DCT data flow*

During clinical trials, data transactions are updated in blocks stored as Hash codes generated against the transactional data [51]. In this paradigm, DCT data transactions follow a similar pattern, with wearable devices facilitating data transformation through the IoT on the blockchain platform. Patients remain connected to IoT devices 24/7, with data automatically recorded on distributed Hyperledger Fabric and active ledger channels accessible by authorized parties [52]. A patient-based Merkle tree is established for country-1's patient-related clinical trial activities within the z-TAB model. Similarly, other countries participating in DCTs adopt similar Merkle tree structures within this model. The patient-based Merkle tree for the depicted country is illustrated in Fig. 11.

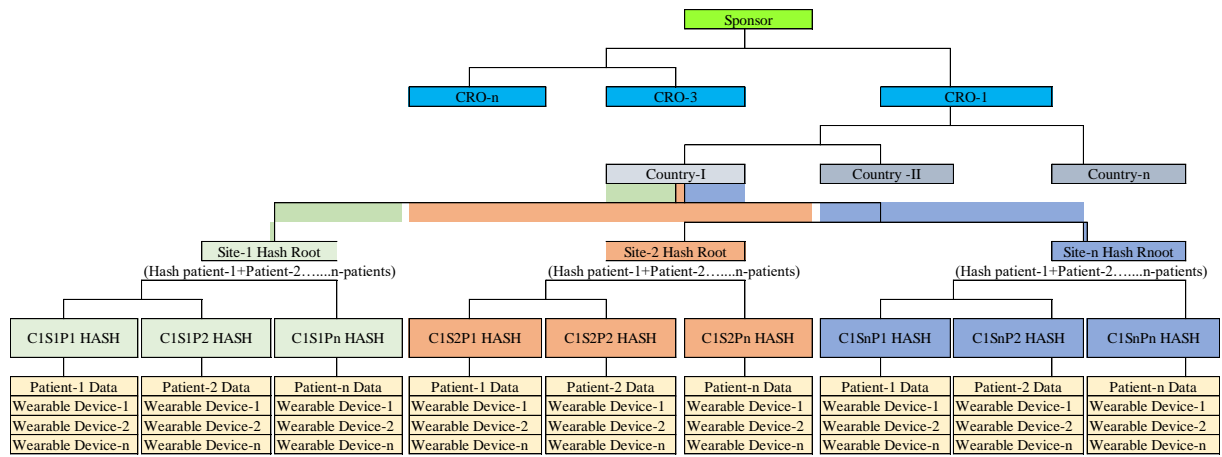

**Fig. 11.** Patient-based Merkle tree of the country.

*4.5. EigenTrust-based practical byzantine fault tolerance (T-PBFT): Consensus protocol*

To enable the addition of new blocks on the blockchain with trust and acceptability during data transactions among all blockchain nodes, consensus protocols are imperative. Several probabilistic consensus algorithms, such as proof of work (PoW), proof of stake (PoS), BFT, and PBFT, are utilized to achieve mutual consensus, trust, and security among decentralized nodes on the blockchain. However, these algorithms have limitations concerning power consumption, efficiency, scalability, and view change issues.

The proposed z-TAB model introduces the T-PBFT consensus algorithm to enhance scalability on a large-scale distributed network of DCT nodes across countries. T-PBFT reduces the probability of the view change process and incorporates group signatures alongside mutual supervision to bolster its robust and resilient application [53]. Eigentrust ensures higher trust values by establishing a trustworthy consensus group, preventing lower trust nodes from participating in the consensus protocol, and enhancing consensus efficiency. This multistage consensus T-PBFT protocol involves evaluating DCT nodes, forming a DCT consensus group, and endorsing the consensus process of all nodes on the blockchain.

The EigenTrust model calculates a unique global trust value for every node in the network by recording the transaction history between nodes. The global trust value can be computed via Eq. (1) as follows:

$$T_i = C_{1i}T_1 + \cdots + C_{ni}T_n \; , \tag{1}$$

Where $T_i$ represents the global trust value of node $i$, and $C_{ij}$ represents the global trust value of node $i$ to node $j$.

The relationship between two nodes is "nodes with transactions and nodes without transactions," as presented in Fig. 12. Based on such transactions, the EigenTrust Model uses three types of trust values during transactions among nodes, which are discussed below.

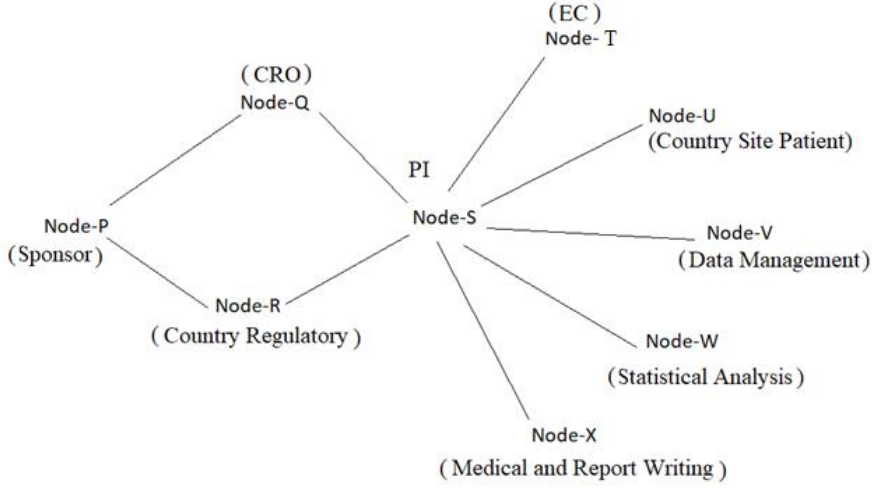

**Fig. 12.** Relation graph of DCT nodes [53].

Direct trust value ($C_{pq}$): It can be evaluated between $node_p$ (sponsor) and $node_q$ (CRO) or $node_p$ (Sponsor) and $node_R$ (Country regulatory) because of direct transactions and can be defined using Eq. (2).

$$S_{pq} = sat_{(node_p, node_q)} unsat_{(node_p, node_q)} \qquad (2)$$

where $sat$ and $unsat$ represent the number of satisfactory and unsatisfactory transactions, respectively, between $node_p$ and $node_q$. $node_p$ and $node_q$ are connected directly where a satisfactory transaction between $node_p$ and $node_q$ is achieved through the rules that nodes need to follow to reach an agreement. The proportion of satisfactory transactions must be higher than the unsatisfactory transactions to measure the direct trust value ($C_{pq}$).

Hence, the direct trust value $C_{pq}$ can be computed through Eq. (3).

$$C_{pq} = \frac{\max(S_{pq}, 0)}{\sum_x \max(S_{px}, 0)} \qquad (3)$$

where $x=q$ and R.

Recommended trust value ($C_{ps}$): The $node_p$ and $node_s$ (PIs) do not conduct any transactions; thus, $C_{ps}$ can be estimated between these two nodes. The basis of the $C_{ps}$ evaluation is transitive trust, and its value is related to the direct trust value. Then the $C_{ps}$ can be represented using equation (4).

$$C_{ps} = \sum_k C_{pk} C_{ks} \qquad (4)$$

where $k = q$ and R.

Global trust value ($T_p^{k+1}$): It is a measurable degree of trust in which a system evaluates nodes. The global trust value of $node_P$ integrates every DCT node trust value in the blockchain network system and adds to each node current global trust value. This value will be the basis of the evaluation index for the trust degree of $node_P$.

Initially, the T-PBFT consensus protocol establishes a global trust for nodes, serving as the foundation for the consensus group. Nodes with high trust values are subsequently selected from this consensus group. As the consensus process unfolds, the number of participating nodes decreases, enhancing the efficiency of T-PBFT in large-scale environments [54].

The global trust value dynamically changes across blocks once a new block is appended to the blockchain and new transactions occur. T-PBFT initiates a new round of the consensus process accordingly. This iterative process continues as transactions progress.

The T-PBFT consensus process is executed in three phases within the z-TAB model.

*4.5.1. Phase-1: Calculation of node trust (direct trust value and recommended trust value)*

The node trust calculation among the network's DCT "*N*" nodes is initiated by directly computing the direct trust value between nodes [26]. We compute the recommended trust value for two nodes where direct transactions do not occur. Then, the global trust can be calculated using these values from Algorithm 1 to Algorithm 4.

**Algorithm 1.** Transaction and no-transaction among nodes on z-TAB model.

Input: $node_i$, Set of Nodes
Output: *TxNodes, NonTxNodes* (based on transaction information)
1. $TxNodes \leftarrow \acute{\O}$,
   $NonTxNodes \leftarrow \acute{\O}$;
2. **For** $node_j \in Nodes$ **Do**
3.     **If** $node_j$ do transaction with $node_i$, **Then**
4.         $TxNodes \leftarrow node_j$
5.     **Else**
6.         $NonTxNodes\ (s,_{T,U,V,W},x \leftarrow p)$
7.     **End**
8. **End**

Algorithm 1 depicts the transaction and no-transaction among nodes on the z-TAB model. $Node_P$ performs the transaction to $Node_Q$ and $Node_R$, so these (*P, Q, R*) are transaction nodes while the other nodes ($s,_{T,U,V,W},x$) are not transacting with the $Node_P$, so these are non-transaction nodes. Algorithm 1 computes the process of determining the direct trust value where nodes are in relationship with the transaction. The direct trust value ($C_{ij}$) is estimated between *i=P* and *j=QRSTUVWX*. It takes $node_i$ and its direct *TxNodes* (QR) as input, then calculates the absolute satisfaction value $S_{ij}$ by analyzing the previous historical node records (in the form of hash values) based on satisfied and unsatisfied transactions. Then, the final direct trust value $C_{ij}$ is calculated between $node_i$ and $node_j$.

**Algorithm 2.** Calculation of *TxNodes* trust/Direct Trust Value

Input: $node_i$, *TxNodes* of $node_i$
Output: Direct trust value $C_{ij}$
1.   $C_{ij} \leftarrow 0$
2.   **For** $node_j \in TxNodes$ **Do**
3.       $S_{ij}= Sat(ij) - unsat(ij)$
4.       $S_{Total}=\sum max(S_{ij},0)$
5.   **End**

6. **If** $S_{Total}=0$, **Then**
7.     Set $C_{ij}= 1/N$, where $N$=Size of nodes
8. **Else**
9.     **For** $node_j \in TxNodes$ **Do**
10.         $C_{ij}= max(S_{ij},0)/S_{total}$
11. **End**

Algorithm 2 estimates the recommended trust value, and it takes $node_i$, *TxNodes* of all nodes and *non TxNodes* where the transaction relationship is not present, as inputs. The direct trust values help establish the transaction pathway. If $node_i$ (P) does not have a direct transaction with $node_j$ (STUVWX), then $node_k \in$ TxNodes needed in which the transaction is completed with target $node_j$ and compute the recommended value to establish the transaction between *non-txNodes*. The value is the product of $C_{ik}$ and $C_{kj}$. If no obstruction in the path exists, then the recommended trust value can be computed iteratively by different transaction paths among *Non TxNodes*.

**Algorithm 3.** Calculation of Non TxNodes trust)/Recommended Trust Value.

Input: $node_i$, *TxNodes, non TxNodes* of $node_i$
Output: Recommended trust value $C_{ij}$
1. $C_{ij} \leftarrow 0$;
2. Determining transaction pathway between $node_i$ and $node_j$ ;
3. **For** $node_j \in$ *NonTxNodes* **Do**
4.     **If** $node_k \in TxNodes$ $node_i$ and $node_k \in$ *non TxNodes* of $node_j$, **then**
5.         $C_{ij}= \sum C_{ik}C_{kj}$ ;
6.     **Else**
7.         Compute $C_{ij}$ ;
    **End**
8. **End**

All nodes establish local trust based on direct and recommended values. The global trust value is required to obtain the node's full trust level. Initially, the trust value of all nodes was $1/N$, where $N$ is the total number of nodes present in the DCT network system. A global trust value is needed when a new block is added to the blockchain network. Algorithm 4 depicts the calculation for the global trust value.

**Algorithm 4.** Calculation of Global trust.

Input: $node_i$, node set *Nodes*
Output: Global trust value of $node_i$
1. $T_i \leftarrow 0$;
2. **For** $node_j \in Nodes$ **Do**
3.     $T_i = \sum C_{ji}T_j$ ;
4. **End**

For $node_i$, its global trust value is the sum of the product of the local trust value and the other node's corresponding global trust value. The global trust value is a dynamic value and is affected by the different network nodes. Such a dynamic evaluation method assists in accurate node trust determination and minimizes the low credit nodes for consensus.

*4.5.2. Phase-2: Building a consensus set among nodes*

The EigenTrust model calculates the overall trustworthiness of nodes within the blockchain network, facilitating the formation of the blockchain consensus group. Instead of including all nodes in the consortium blockchain, only those with higher global trust values are chosen. When a node's global trust exceeds a predefined threshold, it is added to the system nodes, optimizing the efficiency and scalability of the consortium blockchain. These global trust values are dynamic, leading to fluctuations in the composition of blockchain consensus nodes over time.

To overcome fluctuations in global trust in the blockchain and build a trusted environment, a certain percentage of nodes with higher global trust values are selected to construct a consensus group. The steps involved in this process are presented in Algorithm 5.

**Algorithm 5.** ConsensusGroup Construction

Input: node set $Nodes$, Global trust set $T$, A constant percentage of nodes $s$ ($0<s\leq1$)
Output: ConsensusGroup
1. ConsensusGroup ←Ø;
2. **Sort** $Nodes$ by $T$;
3. **For** $node_j \in Nodes$ **Do**
4.    **If** $T_i$ is in the top $s$ **Then**
5.       Add $node_i$ into ConsensusGroup;
6.    **Else**
7.       Exclude $node_i$, from ConsensusGroup;
8.    **End**
9. **End**

In Algorithm 5, an empty "ConsensusGroup" is initiated, and nodes with higher trust values are sorted out. In the constant percentage of nodes $s$ and $node_j$ in the set $Nodes$, if the global trust value of $node_i$ is in the top $s$, $node_i$ will be added to the "***ConsensusGroup***"; otherwise, it would be excluded from the "***ConsensusGroup***". Finally, a group of higher global trust values is determined, and the blockchain consensus group is constructed. Only these nodes are trustworthy nodes that will participate in the consensus process of blockchain to enhance the efficiency of the blockchain consensus process.

*4.5.3. Phase 3: Propagation of the consensus process*

The consensus process among the nodes is established by generating a new block through voting within the "ConsensusGroup". If the primary group fails, Byzantine nodes may behave arbitrarily, potentially causing network failure. The replica nodes, whose expired timers, will detect this and initiate the view change process [26,55]. To avoid such view changes and maintain a consistent consensus process, it is advisable to prevent view changes as much as possible. To manage this, a few nodes with higher trust values are selected from the "ConsensusGroup" to form a primary group, replacing the primary node described in Algorithm 6.

**Algorithm 6**. PrimaryGroup

Input: ConsensusGroup, Certain Percentage $m$ ($0<m\leq1$)
Output: PrimaryGroup
1. PrimaryGroup ←Ø
2. **For** $node_j \in$ ConsensusGroup **Do**
3.    **If** $node_i$ with the global trust value in top $m$ **Then**
4.       Add $node_i$ into PrimaryGroup
5.    **Else**
6.       Exclude $node_i$ from PrimaryGroup
7.    **End**
8. **End**

This primary group accelerates the building, recording, reporting, and conforming of the correctness of the newly generated block. The concept of T-PBFT reduces the risk of the view change process caused by the Byzantine fault. The T-PBFT process is divided into a group process, pre-prepare, prepare, commit, and finishing stages, as depicted in Fig. 13, where $N_1$ is the primary node and the secondary replica nodes are $N_2$ to $N_4$. The $N_5$ node is a node out of the consensus group which would be excluded during the consensus process.

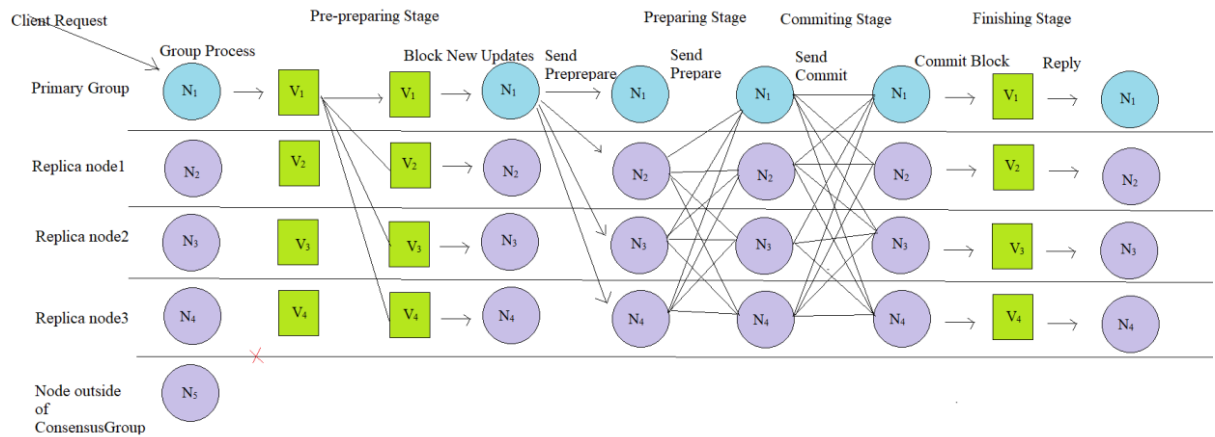

**Fig. 13.** Propagation of the consensus process in T-PBFT [56].

As depicted in Fig. 13, during the group process stage, a node in the primary group transacts the package into a pre-generated block. It broadcasts to the other primary group member nodes for mutual supervision and verification. Once approved, the primary group temporarily stores and records the pre-generated block under the same view. If the primary group fails, another node can immediately replace it to prevent a view change.

In the pre-prepared stage, the primary group will telecast a pre-prepared message along with a pre-generated block and group fingerprint (signature or hash value) to the replica nodes in the consensus group for audit and authentication [57]. The group signature of the consensus group resists view changes during the consensus process. Here, any node can verify the validity of the primary group fingerprint but cannot detect the primary group by which it has been made. In the prepare stage, the replica nodes verify the pre-generated block validity. Every replica node simulates the packaged transaction of the pre-generated block and then computes the block hash (fingerprint of the pre-generated block). If it is consistent with the current block hash, the validation is over, and it is passed. Then, a prepared message will be broadcast to each other with their signatures. Once the number of prepared messages received by the consensus group is greater than $2f$, a reply will be sent to the client, where $f$ represents the Byzantine nodes in the consensus group.

During the pre-prepare stage, the primary group broadcasts a pre-prepare message, including the pre-generated block and the group fingerprint (signature or hash value), to the replica nodes in the consensus group for auditing and authentication [57]. The group signature of the consensus group helps resist view changes during the consensus process. Any node can verify the validity of the primary group fingerprint, although it cannot identify the specific primary group that generated it.

In the prepare stage, the replica nodes verify the validity of the pre-generated block. Each replica node simulates the packaged transaction of the pre-generated block and computes the block hash (fingerprint of the pre-generated block). If this is consistent with the current block hash, the validation is complete, and the block is approved. The nodes then broadcast a prepared message with their signatures to each other. Once the number of prepared messages the consensus group receives exceeds $2f$, where $f$ represents the number of Byzantine nodes in the consensus group, a reply is sent to the client.

In Fig. 13, one node, $N_5$, is out of the consensus process; thus, $f=1$, and $2f=2$. The prepare message is three, which is greater than $2f$ (i.e., 2). The message is then broadcast to the client.

In the committing stage, when the client completes the $f+1$ or more messages of the same reply message from the prepare stage, the pre-generated block is confirmed in the blockchain network, updating their transacting records.

## 5. Zero trust architecture blockchain (z-TAB) model

Zero-trust architecture operates on the principle of "never trust and always verify", treating everything and everyone as untrusted, even within the network. It enforces policies to validate every user or wearable device's activity and promotes a host-based monitoring approach. Integrating zero-trust with blockchain and IoT enhances the system's tamper resistance and prevents unauthorized access.

The proposed z-TAB system ensures data security by leveraging the zero-trust architecture, blockchain, and the Interplanetary File System (IPFS) for data generated by IoT devices, such as wearable devices in DCTs. This system maintains network integration and facilitates efficient communication, reducing the likelihood of real-time attacks through real-time monitoring and policy generation mechanisms.

In this setup, the blockchain ensures DCT data from patients using wearable devices, allowing only recognized nodes to access the network. A dynamic policy mechanism is necessary to create, validate, and identify patients (wearable devices/IoT devices) on blockchain systems. Each node must participate and be authenticated before interacting with other nodes in the system. Blockchain wallets created for each wearable or IoT device help identify, record, and report data transfers automatically using smart contracts, while IPFS stores the encrypted information for further processing [58].

### 5.1. Zone creation on zero-trust architecture

The z-TAB model divides the IoT network into multiple "Zones" based on physical location, priorities, and categories of wearable devices (WDs) from which clinical data are accessed from patients. Similar devices—such as medical earbuds, ECG patches, chest straps, smartwatches, clothing, glasses, helmets, and Oura rings—are grouped into different zones. Clinical data are transferred from these zones according to the wearable devices used by the patients.

For example, Dr. Henri Johnson from America recruited patients (AH1001 to AH...$n$) at their homes. These patients wore various registered wearable devices, and different zones were created for these devices within the z-TAB model (Table 5).

**Table 5**

Zone creation (A-G) for various wearable devices.

| Location | Name of PI | Patient No. | Wearable Devices/IoT | Zones |
|---|---|---|---|---|
| America | Dr Henri Johnson | AHJ1001 AHJ1002 AHJ...*n* | Medical ear bud | A |
| | | AHJ1001 AHJ1002 AHJ...*n* | ECG patch | B |
| | | AHJ1001 AHJ1002 AHJ...*n* | Chest strap | C |
| | | AHJ1001 AHJ1002 AHJ...*n* | Smart watch | D |
| | | AHJ1001 AHJ1002 AHJ...*n* | Clothing | E |
| | | AHJ1001 AHJ1002 AHJ...*n* | Helmet | F |
| | | AHJ1001 AHJ1002 AHJ...*n* | Oura ring | G |

Similarly, at other principal investigator sites, such as Dr. Robert Kole and Dr. Smith, zones are created to collect data from all wearable devices. These zones have Policy Enforcement Points (PEPs) that transfer decisions to the Policy Decision Point (PDP). The PDP accepts or rejects decisions after authentication through encrypted channels on the blockchain [59]. The PDP is interconnected with the Policy Engine (PE) to generate policy access dynamics. When a request is accepted, PEP allows a channel of encryption to facilitate the IoT device (wearable devices and others) interactions.

*5.2. Zero-trust (ZT) architecture on blockchain*

In the zero-trust architecture, no connected devices, systems, or users are trusted by default. Every transaction is monitored and granted only after validation as a legitimate access request. Integrating ZT in the IoT network, particularly with patient wearable devices in DCTs, ensures that all devices are interconnected to provide an immutable environment. The core components of the z-TAB model are presented below in Fig. 14.

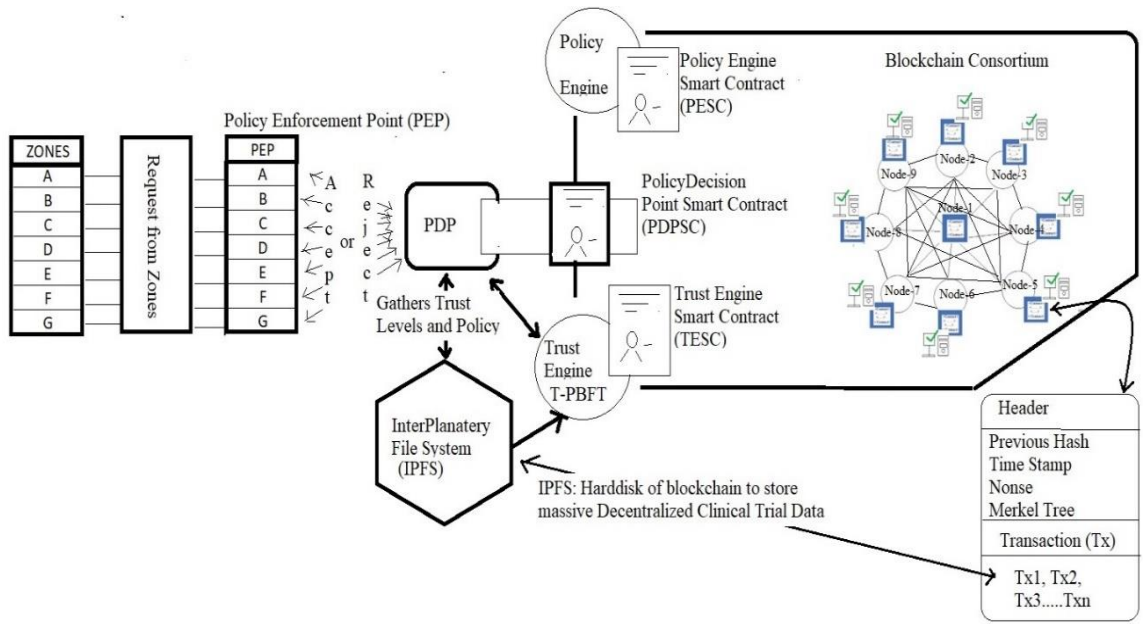

**Fig. 14.** Zero-trust architecture blockchain (z-TAB) model.

A blockchain component is integrated into the zero-trust architecture to facilitate hassle-free communication and data transfer among various IoT devices, enhancing network security and privacy. An Attribute-Based Access Control (ABAC) mechanism is adopted to ensure the security of devices and data management through smart contracts. The PE receives new requests and triggers the policy engine smart contract (PESC) to access new policies for ABAC [60]. In this model, the InterPlanetary File System (IPFS) is used to save attribute-based data received from patients' wearable devices during DCTs. The data transferred to the IPFS are stored as cryptographic hashes in each block, allowing it to be searched and accessed by the generated hash values corresponding to the input data from the patient's wearable devices. Given the large volume of data generated when thousands of patients are enrolled in DCTs across multiple countries, storing all these data directly on the blockchain is impractical. Instead, IPFS is used for off-chain storage, where massive amounts of data can be securely stored using cryptographic hashes. IPFS supports various protocols like File Transfer Protocol (FTP) and the Hypertext Transfer Protocol (HTTPS), and stores information using a distributed hash table, allowing data to be downloaded directly from nodes. This provides greater security and better control over data storage. IPFS allows a secure mechanism for storing clinical data due to automatic resource mapping and hash values (fingerprints of clinical data inputs). It connects to smart contracts, enabling cross-verification of decentralized patient clinical data stored on the IPFS with transactions stored on the blockchain hyperledger. To implement the z-TAB, the trust engine (EigenTrust Byzantine Fault Tolerance; T-PBFT) triggers a trust calculation based on smart contracts and the global trust value of nodes on the blockchain. It calculates the trust level of all wearable devices involved in data transactions by considering the previous data history of each block from different nodes recorded in the ledger.

Finally, the PDP smart contract accepts or rejects IoT/wearable device requests for device-to-device communications in this model.

*5.3. Wearable device registration on blockchain*

Blockchain is a dynamic component of z-TAB, securing transactions among nodes using smart contracts and Hyperledger during decentralized clinical studies. The blocks created on the blockchain are interconnected using cryptographic hashes, providing a secure and immutable environment [61]. When the IoT or wearable devices are registered (Fig. 15), an account is assigned to the connected patient device via smart contracts, initiating the information transaction as hash values. Blockchain wallets ensure the authenticity and transaction anonymity of these IoT devices. T-PBFT is used as a consensus protocol due to the large number of patients in DCTs and the large number of requests. T-PBFT quickly achieves consensus among nodes to add blocks during data transactions. The attributes of IoT devices are stored in the IPFS, and device management smart contracts are installed on IoT devices. Wearable devices on patients record clinical observations (e.g., body temperature, blood pressure, pulse rate, ECG, and other protocol-compliant activities) and transfer communication requests. The PEP acts as an interface, passing requests to the PDP, which triggers the Policy Decision Point Smart Contract (PDPSC). The transaction data are then stored as cryptographic hashes in the distributed hyperledger on the blockchain. If the IoT device is new, the PDPSC generates a new policy and triggers a new Policy Enforcement Smart Contract (PESC), and this transaction is also recorded on the blockchain. When any transaction request is processed, the trust level smart contracts, T-PBFT, are triggered, and a new trust value for the wearable devices is stored as a data transaction on both the blockchain and the IPFS. The clinical data received from all IoT devices associated with patients are linked by cryptographic hashes, trust levels originating from T-PBFT, and policies stored in blocks on the IPFS-based PIP system. The stored clinical data are then used for further validation.

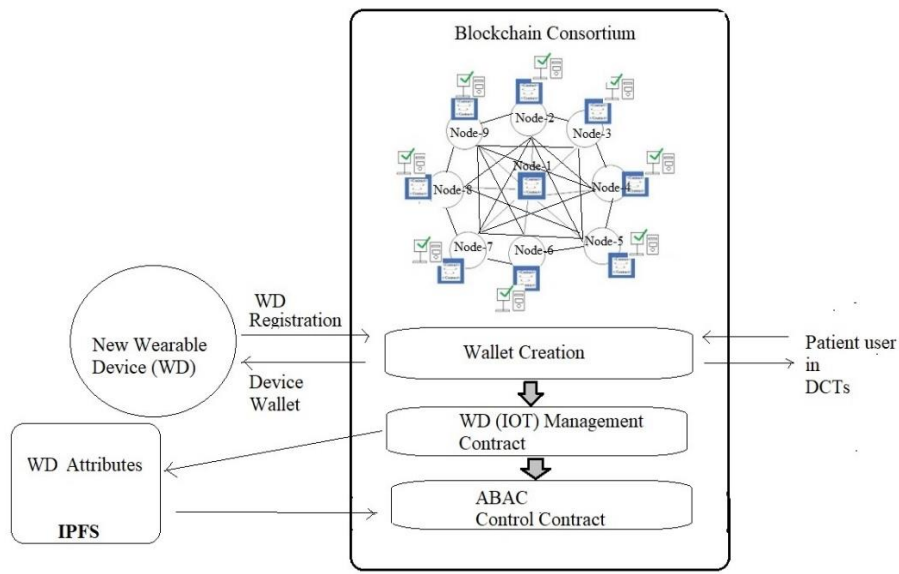

**Fig. 15.** Registration of wearable devices on blockchain.

*5.4. Attribute-based access control (ABAC) mechanism*

The ABAC mechanism on z-TAB approves requests based on the attributes of the sender and receiving nodes. Both sets of attributes form the access control policy, ensuring the security concerns of the receiver's owner (Sponsor). This approach provides the strong dynamics, scalability, and flexibility needed to manage access requests for all wearable device patients use in a wearable device environment. The access control mechanism based on these attributes controls various activities of DCTs. These activities include regulatory approval, the ICF process, EC document submission and approval, clinical site identification, study document and resource availability, patient identification, patient enrollment based on inclusion/exclusion criteria, patient visit-based activities, wearable device data collection, data validation, data freeze, and study close-out [62]. For example, data such as laboratory-based outcomes, blood pressure, heart rate, temperature, movements, ECG records, sounds, humidity, and light are captured directly from patients' wearable devices. To protect the blockchain network on the z-TAB model, only trusted devices are allowed to communicate with the IoT devices on the network, necessitating the implementation of ABAC policies. The implementation of the ABAC policy mechanism on z-TAB is depicted in Fig. 16.

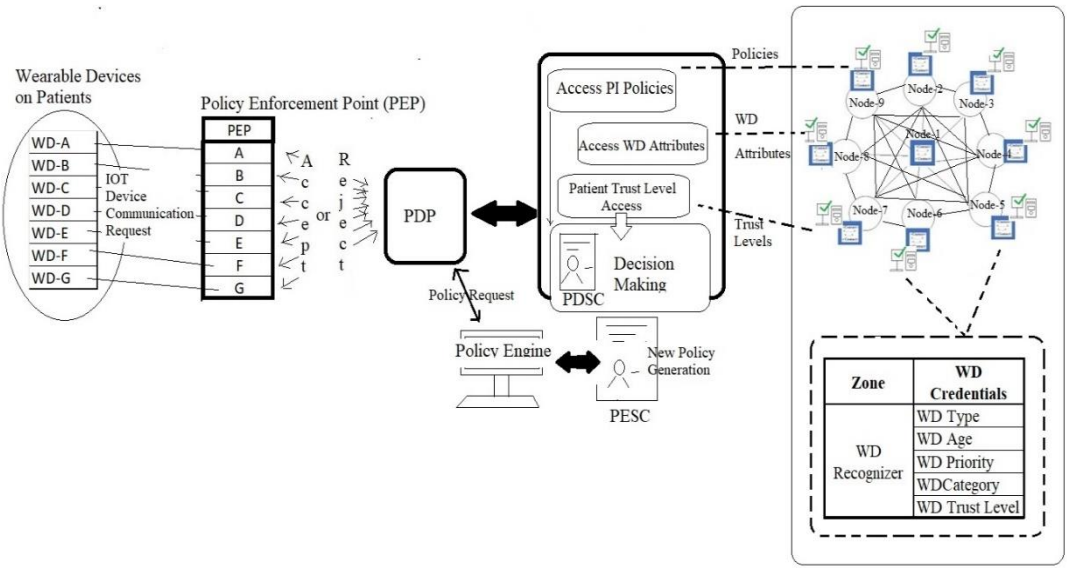

**Fig. 16.** ABAC access policy mechanism implementation in z-TAB.

Within the system, wearable devices (WDs) affixed to patients initiate communication requests with the next node, the Principal Investigator, on the blockchain. These requests are received by the PEP, which then forwards them to the PDP. The PDP retrieves all the attributes recorded from the IoT. The PE determines whether to accept or reject the requests based on zones categorized by device type, category, priority, and trust level. After the PDPSC confirms the authenticity of the request, it establishes a secure encrypted channel for safe device-to-device (D2D) communication [63].

Below, we present the ABAC mechanism, which defines a systematic policy for addressing communication among clinical trial stakeholders.

| ABAC mechanism for communication among the stake holders |
|---|
| **Require:** *Policy* = Patient (WD)$_{attributes}$, PI$_{attributes}$ |
| **Require:** *Patient (WD)$_{attributes}$* = WD recognizer, Type of WD, WD Age, WD Priority, WD Category, WD Zone. |
| **Require:** *PI$_{attributes}$* = WD recognizer, Type of WD, WD Age, WD Priority, WD Category, WD Zone. |
| **Require:** *Environment$_{attributes}$* = Time-stamp |
| **Require:** *TrustLevels* = Patient Trust Level, PI Trust Level, Global Trust Level |
| **if** Permission == 1 **then** |
| AccessGranted |
| **else if** Permission == 0 **then** |
| AccessDenied |

*5.5. Hashed storage of wearable device data through the IPFS*

IPFS is employed within the model to accommodate the vast amount of clinical data generated during DCTs [64]. It serves as a repository for attributes originating from all connected Internet of Things (IoT) devices, smart contracts, and transaction history and ensures data security. Clinical data, including text, audio, video, and images generated by the connected IoT devices, undergo encryption via hash algorithms before being stored in blocks across blockchain nodes (Node-1 to Node-9). Policies and trust levels stored in IPFS are validated against IPFS hash blocks through blockchain transactions, guaranteeing the integrity and non-tampering of stored data and policies [38]. A way of data storage in a blockchain-based zero-trust architecture model is presented in Fig. 17.

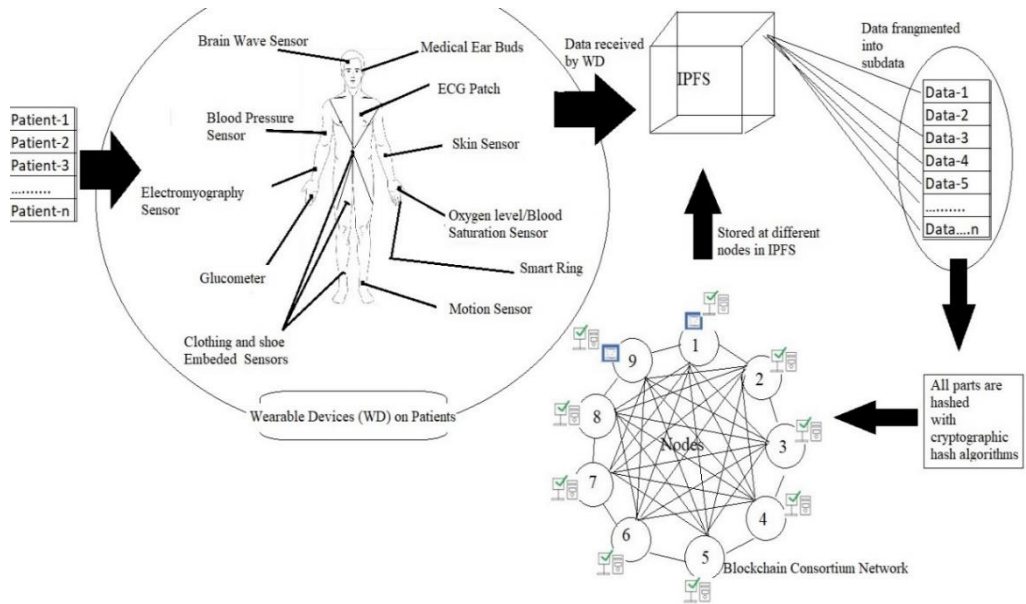

**Fig. 17.** Storage of wearable device data through IPFS

Patients' data received through wearable devices are fragmented into subdata (data-1, data-2…., data-*n*). The cryptographic hashing technique Secure Hash Algorithm-256 is used to generate the hash values of each dataset (Fig. 21). The transaction data are updated on different nodes (Node-1 to Node-9) on the blockchain and stored in the IPFS.

*5.6. Policy decision point (PDP) and policy enforcement*

The PDP has multiple PEPs from which all requests are submitted to the PDP, as shown in Fig. 18. The PDP assesses the policies and device attributes of the IPFS to ensure the current trust level of each IoT device from the Trust engine (T-PBFT). The requests may be accepted or rejected based on authenticity by putting the acquired transacted data depicting the run-time status of the network and other involved IoT devices into policy. If the present request does not suit the policy, the PE generates a new policy for the current scenario, and subsequently, the PESC is triggered.

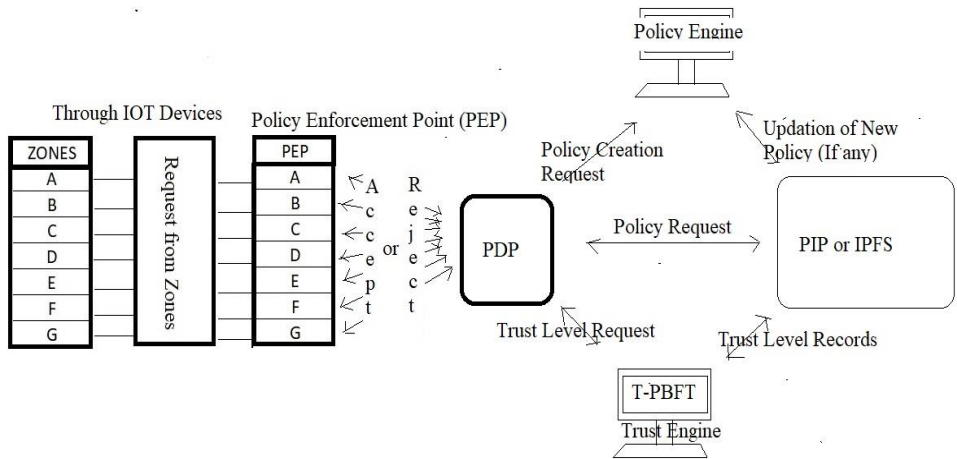

**Fig. 18.** Process of policy decision points (PDPs) and policy enforcement points (PEPs).

*5.7. Trust engine (TE)*

The trust engine is an important component of z-TAB, and it assists in the calculation of IoT devices trust levels in the network. TE is interlinked with the PDP to provide the updated trust levels of patient (wearable device) data transactions to the Principal Investigator (Receiver) IoT devices from policy evaluation [65].

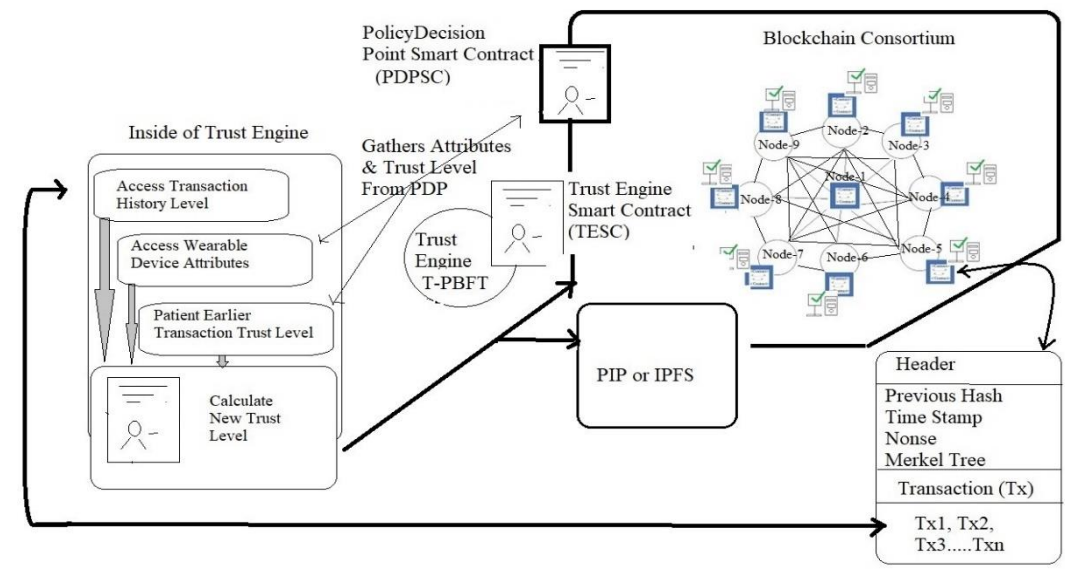

**Fig. 19.** Trust engine on blockchain.

The historical transactions of data are represented by hash values corresponding to patient input data during the execution of clinical trials in a decentralized manner. The PDPSC and T-PBFT function as trust engines to access attributes from wearable devices (which record clinical data from patients). After each node on the blockchain completes verification of the transaction's authenticity and achieves consensus, a new trust level is established.

## 6. Applicability of the z-TAB model for DCTs

The z-TAB model, designed for DCTs worldwide, facilitates the gathering of authentic data on a blockchain platform. Clinical trial sites operate with ethics committees (ECs), which approve of study protocols specific to each site once regulatory bodies in respective countries (such as Country-1: America, Country-2: Brazil, and others) grant clinical study approval. This approval enables Principal Investigators to identify, screen, and enroll patients. Patients were selected for the study based on compliance with the inclusion and exclusion criteria. They utilize wearable devices integrated into the z-TAB system. Access to patient-related activities is controlled via z-TAB's ABAC mechanism, which is managed through registered devices on the IoT network. Patients from whom data are collected register through their wearable devices on the blockchain network, obtaining a wallet with public and private keys. This method establishes a unique patient identification system based on newly registered devices for specific clinical trial activities, ensuring anonymous communication and data security. All communications between nodes (1–9) are encrypted using the SHA-256 algorithm, providing comprehensive protection against unauthorized access.

The entire workflow of z-TAB is outlined, with a focus on two blockchain nodes (Node-6: patient and Node-5: Principal Investigator) in the described steps using the model.

1. Once the wearable device has registered, it becomes part of the blockchain consortium and IoT network, where it can request access to the system.
2. The patients' requests are received by the PEP and directed to the PDP.
3. The PDP gets the attributes and trust level from the PIP, where the DCT activity-based policy is verified by the patient.
4. If the policy exists, the smart contract processes the communication further, and the PDPSC is triggered to accept or reject the request.
5. If a policy is not found, then a policy generation request is made to PE. PESCs started to generate new policies based on the patient activity-related attributes and the trust level, type, and category of wearable device in compliance with the trust level, type, and category of the PI.
6. Once the policy is framed for the attribute, the PEP initiates its enforcement. If access is permitted, PEP signals an encrypted channel on the blockchain consortium to facilitate secure and protected data communication between the patient (Node-6) and the principal investigator (Node-5). The patient is updated on the rejection of the request if it is denied by the PI.
7. The data transaction based on the attributes are stored in PIP where the trust level of wearable devices and PI attributes are verified. The requests and decisions taken are stored in the blockchain-based Distributed Ledger System (DLS) in the form of hash values (Table 5), making the system more immutable on the IoT network. The malicious attack or alternation in PIP can be easily detected by matching the records in a DLS.
8. In the end, the Trust Engine Smart Contract (TESC) is triggered at every data transaction, and the device's acceptance or denial is updated depending upon this new transaction and the device's previous behavior or hash values.

The applicability of z-TAB through the attribute-based smart contract and trust level leads to implementing policies or generating policies where an attribute policy is not found, as shown in Fig. 20, along with a brief description of the steps.

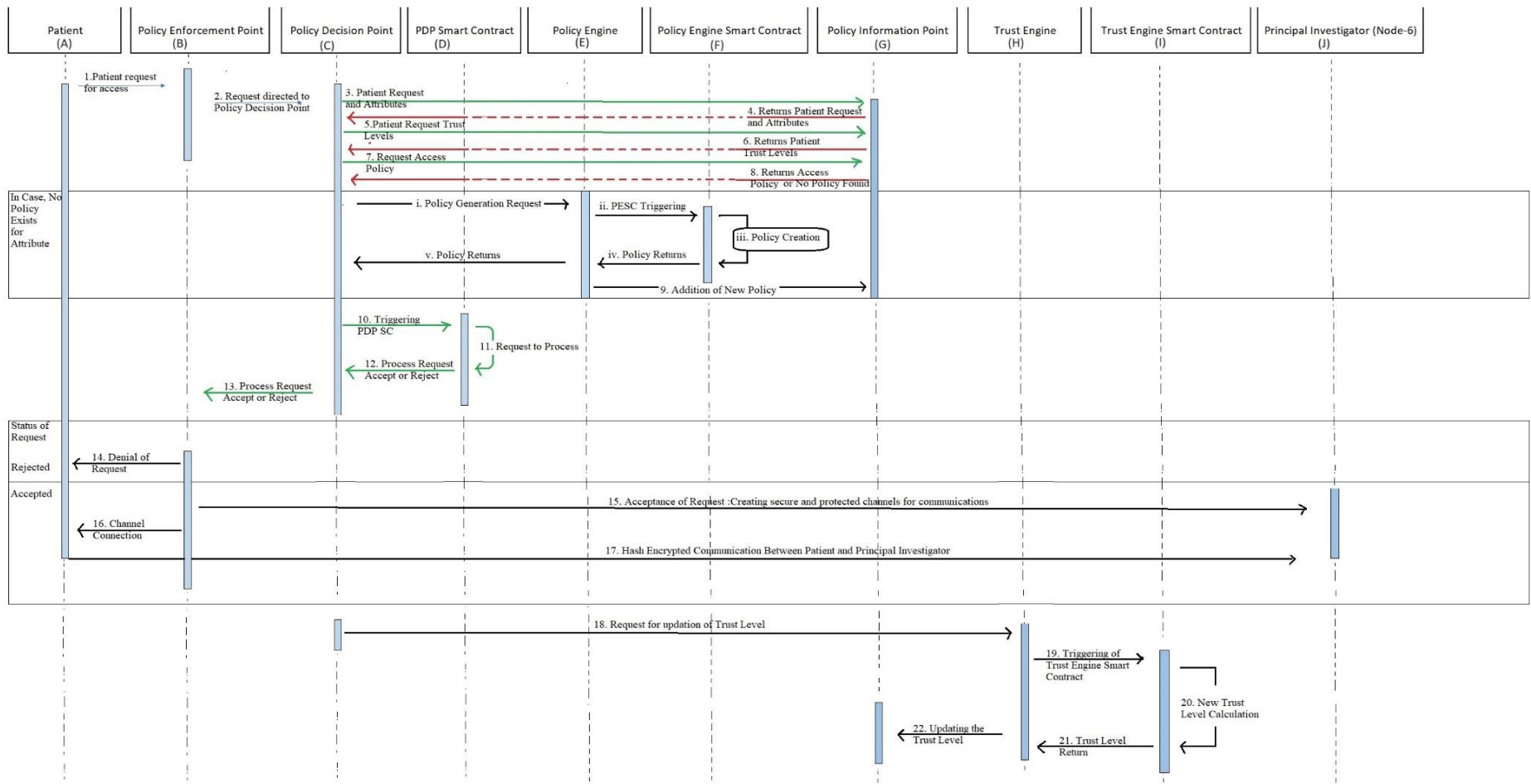

**Fig. 20.** Sequence diagram of z-TAB applicability and policy enhancement (A: Patient; B: Policy Enforcement Point; C: Policy Decision Point; D: PDP Smart Contract; E: Policy Engine; F: Policy Engine Smart Contract; G: Policy Engine Point; H: Trust Engine; I: Trust Engine Smart Contract; and J: Principal Investigator)

*6.1. Functioning of the z-TAB Model and Policy Enforcement*

The functions of the z-TAB model and policy enforcement are depicted in Figure 20. First, the patient requests access to the network; then, it will be directed to the PEP. The request is forwarded to the PDP, patient request and attributes, and trust level request access policy received by the PIP. If a policy is not found, then a policy generation request will be sent for PESC triggering, a new policy creation process will be started, a new policy will be created, and the policy will return to the PDP, directing it to the PIP. As a result, a new policy is added, resulting in triggering of the PDP smart contract.

A request to process the activity will be sent where the process request will be accepted/rejected by PDP forwarding to PEP. When the request is accepted, it creates a secure and protected channel for communications in the form of hash-encrypted communications between the patient and the principal investigator. A request for updating the trust level from the PDP to the TE is passed, and the TESC receives the updates. Thus, a new trust level calculation is achieved, and the trust level returns to TE and is updated for the trust level to PIP.

*Case 1: Demonstrating the applicability of the zero-trust architecture blockchain (ZTAB) model for DCTs.*

Consider an American clinical study setting with three investigators tasked with patient recruitment, highlighting the model's relevance. Dr. Henri Johnson from a hospital (serving as a clinical study site) in America collaborates with Dr. Robert Kol and Dr. Smith, who are also recruiting patients. The registration of coordinators, patients, PIs, and other stakeholders is completed using z-TAB procedures. Patient-related activities are overseen by study coordinators designated by the PIs at clinical study sites, with clinical data transferred from various patient wearable devices (across different Zones A to G) in the form of Hash values (as shown in Table 6) on the blockchain platform.

**Table 6**
Zonewise hash values of patients during DCT implementation

| Country No. | Location | Name of Site | Patient No. | Zones | Cryptographic Hash Function | Hash Value |
|---|---|---|---|---|---|---|
| I | America | Dr Henri Johnson | AHJ1001 | "A" to "G" | SHA-256 | 6357526aeb58f22724f28ff188212d2a14d0c1b090c3a738b025c812712a4a3c |
| | | | AHJ1002 | | | 834192d2491d77daa92deeadf40786c8a550cb96d0c434bec5a2624133e483c5 |
| | | | AHJ...*n* | | | 444067ec31f6d3c30abad17badbaf88dc3b63c82ed2790ba129ab6261e6a05b0 |
| | | Dr. Robert Kol | ARK1001 | "A" to "G" | | b173feee7b0e68aefa50da3c5c0e0189f0148dfed967f3f2709ffb65aeae470a |
| | | | AHR1002 | | | efb66544ed096695c6a99d21d9e09e6c18fb9b53526cb018f996fe053eadc58c |
| | | | AHR...*n* | | | cb242af0f3f0410abf8d5a3cfaa5533ef4b63fcdac3838cf3b0f8b50a8fbca80 |
| | | Dr. Smith | AS1001 | "A" to | | c968ab9acbdee4054f2c78ecdd4b7f7635ff464ef5967f5371237bedd6ea0b27 |
| | | | AS1002 | | | ba18b970dca5e30cf33bf99242c54114f1c4e8c1e192d5695b3b0b014cbee790 |

| | | |
|---|---|---|
| AS...*n* | "G" | 7d642d6db6fc08668bb6dcb4f6436916d0d76b0c7c2465ea5cc153d80c5c2c15 |

Every patient is wearing wearable devices (IoTs) as per the clinical study requirements, and various Zone-A, B, C, D, E, F, Z are assigned to these WDs, such as medical earbuds, ECG patches, chest straps, smart watches, clothing, helmets, and Oura rings. Apart from this access to digital data, additional devices may also be registered as per the clinical study requirements for z-TAB.

The clinical trial activities (regulatory approvals, EC approvals, patient identification/screening, enrollment per inclusion–exclusion criteria, site initiation, compliance with IMP storage, IMP dispensing, and administration, laboratory test assessments, study visit assessments, monitoring observations, data clarifications, and close-out visits) are captured through mobile cloud computing. The clinical data of individual patients (1001----*n*) from individual sites of different countries are transferred directly on the blockchain through wearable devices in the form of hash values via the secure hash algorithm-256, which produces 64-character hash values in hexadecimal form (0-9, *a-f*). The hashes of individual patients from different wearable devices are depicted in Fig. 21.

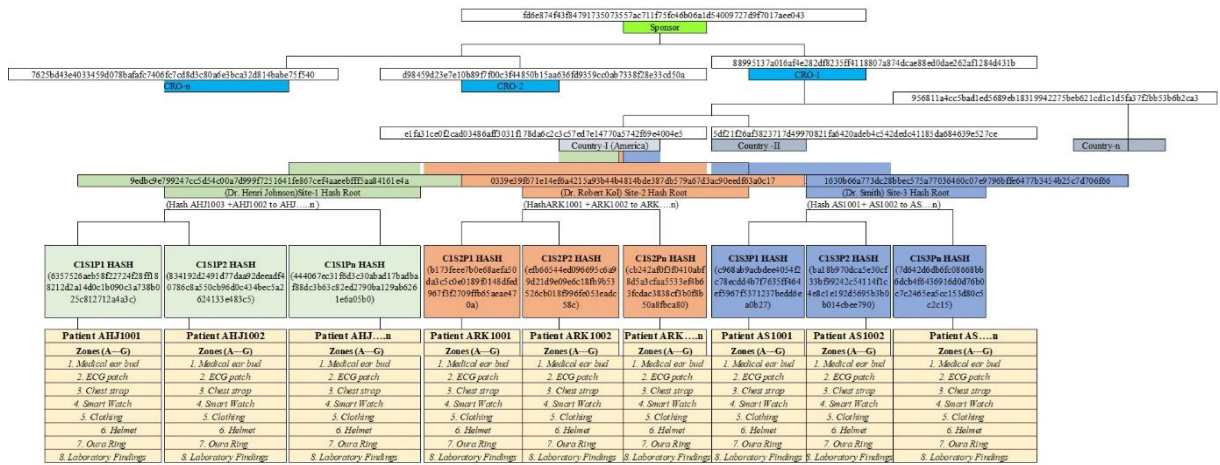

**Fig. 21.** Merkle tree data structure (C: Country, S: Site, P: Patient (Example: C1S1P1, means in country 1 site 1 and patient number 1, C1S1P2, means in country 1 site 1 and patient number 2…... so on)).

The z-TAB applicability initiates from wearable devices requesting zone formation, and the PEP creates zones that may be accepted or rejected based upon the authenticity of patient devices during DCTs. The PDP is automatically executed by PDPSCs, and the PE decides to accept or reject the request. The data transacted on the blockchain consortium via the patients' wearable devices are stored through the IPFS, as shown in Fig. 22.

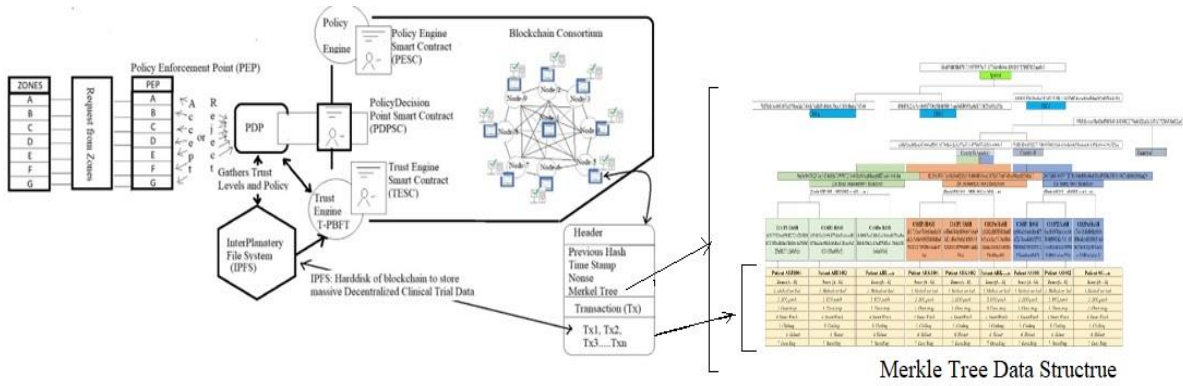

**Fig. 22.** Blockchain, wearable devices, and data storage on IPFS.

All nodes (node-1 to node-9 of the blockchain) are part of T-PBFT (operates under the TESC) to ensure the mutual consensus of the nodes. The Merkel tree data structure shown in Fig. 21 indicates the individual patient's data transfer through the IoT (Wearable devices) hash values, which are ultimately converted into a single hash of sponsor, as shown below:

"fd6e874f43f84791735073557ac711f75fc46b06a1d54009727d9f7017aee043". It contains all patients' clinical data from America, which includes clinical study sites (Dr. Henri Johnson, Dr. Robert Kol, Dr. Smith). Principal Investigator (PI) Dr. Henri Johnson (AHJ1001, AHJ1002…. AHJ….n), Dr. Robert Kol (ARK1001, ARK1002, and ARK….n) and Dr. Smith (AS1001, AS1002, and AS….n) recruited patients. The patient data transaction using the SHA-256 transforms into data signatures in the form of hash values and constructs a Merkle tree data structure. Every PI has its own hash value against the receiving data text (Zones A-G) from the recruited patients (Table 7).

**Table 7**
Hash values of principal investigators through the Merkel tree data structure.

| Principal investigator's hash values | | Patient Hashes |
|---|---|---|
| Dr Henri Johnson | 9edbc9e799247cc5d54c00a7d999f7251641fe867cef4aaeebfff5aa84161e4a | +C1S1P1 +C1S1P2 +C1S1Pn |
| Dr. Robert Kole | 0339e39f671e14ef6a4215a93b44b4814bde387db579a67d3ac90eedf63a0c17 | C1S2P1 +C1S2P2 +C1S2Pn |
| Dr. Smith | 1630b66a773dc28bbec575a77036460c07e9796bffe6477b3454b25c7d706f66 | +C1S3P1 +C1S3P2 +C1S3Pn |

These principal investigators' signatures (Hash Values) are forming a country data signature of America that is e1fa31ce0f2cad03486aff3031f178da6c2c3c57ed7e14770a5742f69e4004e5. Similarly, other country signatures are as follows:

Country-II: 5df21f26af3823717d49970821fa6420adeb4c542dedc41185da684639e527ce

Country-n: 956811a4cc5bad1ed5689eb18319942275beb621cd1c1d5fa37f2bb53b6b2ca3, and these signatures form one signature of

CRO-1: 88995137a016af4e282df8235ff4118807a874dcae88ed0dae262af1284d431b and other CRO signatures are as follows:

CRO-2: d98459d23e7e10b89f7f00c3f44850b15aa636fd9359cc0ab7338f28e33cd50a

CRO-n: 7625bd43e4033459d078bafafc7406fc7cd8d3c80a6e3bca32d814babe75f540

The embedded data transactions from these CROs will be under the sponsor, and the sponsor will access the clinical trial data through CRO signatures, which represent the Sponsor signature:

Sponsor: fd6e874f43f84791735073557ac711f75fc46b06a1d54009727d9f7017aee043

The entire patient details and data flow are monitored directly by the CRO and sponsor via the concerned principal investigators of the respective sites in various countries.

## 7. z-TAB Model Evaluation in the Operation Management of Decentralized Clinical Trials

In DCTs, the IoT devices, such as patients' wearable devices, operate seamlessly without human intervention throughout their data transaction processes. In DCT scenarios, millions of these devices and sensors interconnect to form a smart contract network facilitating smart data transactions among stakeholders (Node-1 to Node-9). Primarily focused on patients (Node-6) and Principal Investigators (Node-9), DCT involves creating, recording, correcting, verifying, reporting, and archiving data in accordance with approved study protocols. Any vulnerability in data security not only hampers the drug approval process but also jeopardizes the lives of thousands of patients awaiting new treatment interventions. The proposed z-TAB model not only ensures the authentication and authorization of stakeholders on z-TAB, as demonstrated in the patient and principal investigator use case (depicted in Fig. 20) but also safeguards the confidentiality and privacy of generated clinical trial data within the blockchain consortium. Evaluating the zero-trust architecture blockchain model on parameters such as immutability, authenticity, privacy, supply chain parameters (such as temperature and humidity for biological samples), mutual consensus, and transparency in data transactions among stakeholders ensures the practical operability of the model during the execution of clinical trials on the DCT framework [40].

### 7.1. Immutability

The clinical trial data are collected virtually through patients' wearable devices (IoTs) and encrypted as a hash in a block. The data transaction of the individual block and the previous block's hash generate the hash using the SHA256 algorithm for the current block header, which is carried forward to the next block. Block header hash: SHA256[SHA256(previous block hash + Time Stamp+ Merkle root+ nonce)]

It's a practicality of blockchain where blocks are interconnected in a chain. SHA256 algorithm generates a 64-string irrespective of the length of data inputs through the wearable devices of individual patients, as shown in Fig. 23.

| Block Header | Patient AHJ1001 | | Block Header | Patient AHJ1002 | | Block Header | Patient AHJ….n | |
|---|---|---|---|---|---|---|---|---|
| | Previous Hash | Nonce | | 6357526aeb58f22724f28ff188212d2a14d0c1b090c3a738b025c812712a4a3c | Nonce | | 834192d2491d77daa92deeadf40786c8a550cb96d0c434bec5a2624133e483c5 | Nonce |
| | Time Stamp | Merkle Root | | Time Stamp | Merkle Root | | Time Stamp | Merkle Root |
| Body | Data Transactions | | Body | Data Transactions | | Body | Data Transactions | |
| | WD-A | TX1 | | WD-A | TX1 | | WD-A | TX1 |
| | WD-B | TX2 | | WD-B | TX2 | | WD-B | TX2 |
| | WD-C | TX3 | | WD-C | TX3 | | WD-C | TX3 |
| | WD-D | TX4 | | WD-D | TX4 | | WD-D | TX4 |
| | WD-E | TX5 | | WD-E | TX5 | | WD-E | TX5 |
| | WD-F | TX6 | | WD-F | TX6 | | WD-F | TX6 |
| | WD-G | TX7 | | WD-G | TX7 | | WD-G | TX7 |
| SHA256-Hash | 6357526aeb58f22724f28ff188212d2a14d0c1b090c3a738b025c812712a4a3c | | SHA256-Hash | 834192d2491d77daa92deeadf40786c8a550cb96d0c434bec5a2624133e483c5 | | SHA256-Hash | 444067ec31f6d3c30abad17badbaf88dc3b63c82ed2790ba129ab6261e6a05b0 | |

**Fig. 23.** Hash values against the WD-generated DCT Data.

The model functions on the principle of blockchain and copies every transaction recorded in any retrospective unauthentic modification (addition/deletion/alteration) that occurs in recorded data (TX1–TX7) of any patient; the modification affects the respective Merkel root, resulting in a change in the current block header hash. Thus, ultimately disrupting the chain of blocks on the blockchain consortium. Here, T-PBFT also protects the unauthorized node or action, as it would not have been part of any of three values ($C_{pq}$, $C_{ps}$, and $T_{pk+1}$) and would not have been allowed to be added in the blockchain because it changes the entire hash of subsequent blocks.

| Block Header | Patient AHJ1001 | | Block Header | Patient AHJ1002 | | Block Header | Patient AHJ….n | |
|---|---|---|---|---|---|---|---|---|
| | Previous Hash | Nonce | | 49a8d9eec955101be1b30dc5ef8e6a337fc3fec4017998e28f6cba24b1ca980d | Nonce | | 08a393541c02fe8c621fd7e27fbd29591a8894cdf800ea132e158931814ad551 | Nonce |
| | Time Stamp | Merkle Root | | Time Stamp | Merkle Root | | Time Stamp | Merkle Root |
| Body | Data Transactions | | Body | Data Transactions | | Body | Data Transactions | |
| | WD-A | Tx11 | | WD-A | TX1 | | WD-A | TX1 |
| | WD-B | TX2 | | WD-B | TX2 | | WD-B | TX2 |
| | WD-C | TX3 | | WD-C | TX3 | | WD-C | TX3 |
| | WD-D | TX4 | | WD-D | TX4 | | WD-D | TX4 |
| | WD-E | TX5 | | WD-E | TX5 | | WD-E | TX5 |
| | WD-F | TX6 | | WD-F | TX6 | | WD-F | TX6 |
| | WD-G | TX7 | | WD-G | TX7 | | WD-G | TX7 |
| New Changed-Hash | 49a8d9eec955101be1b30dc5ef8e6a337fc3fec4017998e28f6cba24b1ca980d | | New Changed-Hash | 08a393541c02fe8c621fd7e27fbd29591a8894cdf800ea132e158931814ad551 | | New Changed-Hash | 93508ad34fb695212143bd50e3ce36b7929f622c2cb8a6faf6360d053f849c73 | |

**Fig. 24.** Changed hash values against changed DCT data (TX11).

Here, it has been observed that once the change in transaction data "TX1" changed to "Tx11", the Merkle root changes completely because the Merkle root is part of a block, and such changes may result in change of entire Merkle root. Patient AHJ1001 recorded the data transaction (TX1 to TX7), which has generated the hash of "6357526aeb58f22724f28ff188212d2a14d0c1b090c3a738b025c812712a4a3c". Other patients AHJ1002 and AHJ…*n* generated the hashes "834192d2491d77daa92deeadf40786c8a550cb96d0c434bec5a2624133e483c5" and "444067ec31f6d3c30abad17badbaf88dc3b63c82ed2790ba129ab6261e6a05b0", respectively. However, hashing algorithms are deterministic, resulting in a different output if the input transaction data change. Therefore, a change in data may change the Merkle root and lead to the hash of the next proceeding blocks on the blockchain. Here, a small change in Patient AHJ1001 recorded the data transaction (TX1 to TX11), resulting in a completely new changed hash as "49a8d9eec955101be1b30dc5ef8e6a337fc3fec4017998e28f6cba24b1ca980d".

Other patients AHJ1002 and AHJ….n hash will also be changed to "08a393541c02fe8c621fd7e27fbd29591a8894cdf800ea132e158931814ad551" and "93508ad34fb695212143bd50e3ce36b7929f622c2cb8a6faf6360d053f849c73" respectively. Thus, the blocks next to it will no longer have links, as the previous hash will not match the new block. As a result, any broken link between two blocks will make the block invalid or unauthorized. All nodes cannot mutually accept such modifications because each node has the previous copies of the data transaction as a "hash." T-PBFT never endorses the unauthenticated data transaction and rejects the transaction. Thus, model z-TAB ensures immutability during DCT data transactions across patients and other stakeholders.

### 7.2. Privacy and security

Ensuring the privacy and security of patients' clinical data entails retaining control over how users' data are collected and managed within the z-TAB model. All the transactions among nodes are routed through Hyperledger Fabric on the blockchain. This ledger enables active channel nodes to share clinical trial data while restricting access for other nodes. Specifically, the data transaction copy resides solely with active channel nodes, such as CRO (2), EC (4), PI (5), and patients (6), within the patient enrollment channel shown in Table 8. In contrast, inactive channel nodes such as Sponsor (1), Regulatory (3), Data Management (7), Statistical Analysis (8), and Report Writing (9) do not possess copies of patient data. This setup ensures that privacy among active channel nodes remains protected, as data are not divulged to other nodes that are not part of the channel within the Hyperledger Fabric system on the blockchain, as depicted in Fig. 3.

**Table 8**
Privacy and security in the patient enrollment channel.

| Name of Channel | Nodes of private channel | Active and inactive nodes on private channel | Data transaction copy |
| --- | --- | --- | --- |
| Patient enrollment channel | Node-4,5,6 | Active channel: 4,5,6 Inactive:1,2,3,7,8,9 | Only active channel nodes will have the same data transaction copy, and other inactive will not have the information |

The Merkel tree structure facilitates data transactions among active nodes by storing authentication credentials for Node 6 (patients), Node 5 (PIs), and Node 4 (ECs), along with the ID numbers of patients' wearable devices, on cloud computing. Patient clinical data information flows through this tree structure exclusively within the active nodes of private channels, safeguarding data privacy and ensuring clinical data security.

In the security-focused z-TAB model, a malicious node attempting to breach the blockchain nodes must falsify all authentication credentials within the Merkel tree of active nodes to obtain the same Merkel root as genuine nodes. However, these authentication credentials are stored as hash values within the blockchain Merkel tree credentials, making it impossible to access or steal them, thus protecting data privacy. Consequently, the z-TAB model can effectively withstand threats from malicious entry or unauthorized attacks.

### 7.3. Mutual consensus

The z-TAB model operates on a blockchain framework without a central authority to validate transactions. Instead, the T-PBFT consensus protocol facilitates mutual consensus among operational nodes within DCTs. This protocol ensures that all active nodes, spanning from Node-1 to Node-9, are informed about the current state of the distributed Hyperledger Fabric system. Doing so enhances reliability and authenticity within the distributed computing environment.

| **Algorithm 7**. z-TAB model on Patient Enrollment Channel |
|---|
| Input: $Node_6$, Set Nodes (1 to 9) |
| Output: $TxNodes(4,5,6)$, $NonTxNodes$ (1,2,3,7,8,9) |
| 1      $TxNodes \leftarrow \acute{Ø}(5 \leftarrow 6; 4 \leftarrow 6)$ |
|       $NonTxNodes \leftarrow \acute{Ø}(1,2,3,7,8,9 \leftarrow 6)$ |
| 2     **For** $node_j \in Nodes$ **Do** |
| 3         **If** $node_j$ (4 and 5) do transaction with $node_i$ (6) **Then** |
| 4             $TxNodes \leftarrow node_j$ (4 and 5) |
| 5         **Else** |
| 6             $NonTxNodes$ (1,2,3,7,8,9 $\leftarrow$ 6) |
| 7         **End** |
| 8     **End** |

The T-PBFT consensus protocol employs Algorithm 7 to evaluate the trust among the nodes (Node-1 to Node-9) within the z-TAB model. This algorithm distinguishes between transactional (*TxNodes*) and non-transactional nodes *(NonTxNodes)*, particularly within the z-TAB model on the patient enrollment channel (Algorithm 7), where transactions occur.

While non-transacting nodes 1, 2, 3, 7, 8, 9 do not participate in transactions among the N nodes (1–9), where the transacting nodes 4, 5, 6 carry out the transaction on the patient enrollment channel. First, Node-6 performs a transaction with Node-4 and Node-5, making these Nodes (6,5,4) the transaction nodes on the Hyperledger fabric. Nodes 1, 2, 3, 7, 8, and 9 do not perform a transaction with Node-6, as shown in Table 9.

**Table 9**
Mutual Consensus on Patient Enrollment Channel Nodes.

| Name of Channel | Nodes of the private channel | Name of nodes | Active and inactive nodes on a private channel | Transacting and non-transacting node |
|---|---|---|---|---|
| Patient enrollment channel | Node-6 Node-5 Node-4 Node-2 | Patient PI EC CRO | Active: 4,5,6 Inactive:1,2,3,7,8,9 | Transacting nodes: 4,5,6 Non-transacting nodes:1,2,3,7,8,9 |

| **Algorithm 8**. z-TAB model on Patient Enrollment Channel |
|---|
| Input: $node_i$, $TxNodes$ of $node_i$ |
| Output: Direct trust value $C_{ij}$ |
| 1.   $C_{ij} \leftarrow 0$, (i=6; j=1,2,3,4,5,7,8,9) |
| 2.   **For** $node_j \in TxNodes$ **Do** |
| 3.       $S_{ij} = Sat(6,5,4) - unsat(1,2,3,7,8,9)$ |
| 4.       $S_{Total} = \sum max(S_{ij}, 0)$ |
| 5.   **End** |
| 6.   **If** $S_{Total} = 0$, then |
| 7.       Set $C_{ij} = 1/N$, where $N$=Size of nodes |
| 8.   *Else* |
| 9.       **For** $node_j \in TxNodes$ **Do** |
| 10.      $C_{ij} = max(S_{ij}, 0) / S_{total}$ |

11. **End**

The direct trust value of the nodes participating in direct transaction relationships is determined by Algorithm 8. The predicted direct trust value ($C_{ij}$) ranges from i=6 to j=1,2,3,4,5,7,8,9. It analyses previous historical node records (hash values) from all nodes based on satisfied and unsatisfied transactions and then calculates the absolute satisfaction value $S_{ij}$ (node-1 to node-9) using $node_i$ (node-6) and its direct TxNodes (node-5 and node-4). The ultimate direct trust value $C_{ij}$ between $node_i$ and $node_j$ was then determined.

**Algorithm 9**. z-TAB model on Patient Enrollment Channel

Input: $node_i$, *TxNodes, non TxNodes* of $node_i$
Output: Recommended trust value $C_{ij}$
1. $C_{ij} \leftarrow 0$; (i=6; j=1,2,3,4,5,7,8,9)
2. Determining transaction pathway between $node_i$ (6) and $node_j$ (1,2,3,4,5,7,8,9)
3. **For** $node_j$, € *NonTxNodes* (1,2,3,7,8,9) **Do**
4.    **If** ($node_k$ (4,5) ∈ *TxNodes* $node_i$ (6)) *and* ($node_k$ (4,5) ∈ *NonTxNodes* of $node_j$, (1,2,3,7,8,9)) **Then**
5.       $C_{ij} = \sum C_{6,4} C_{4,2}$
6.    **Else**
7.       Compute $C_{6,2}$
8.    **End**
9. **End**

Algorithm 9 determines the suggested trust value using $node_i$ (node-6, patient), all nodes' TxNodes, and non-TxNodes (nodes without a transaction relationship). The transaction pathway is established with the aid of direct trust values. The $node_k$ € TxNodes needed in which transaction completed with target $node_j$, computed the suggested value to establish the transaction between non TxNodes (1,2,3,7,8,9) when the $node_i$ (node-6, patient) does not have the direct transaction with $node_j$ (1,2,3,7,8,9). The sum of $C_{ik}$ and $C_{kj}$ yields the value. The recommended trust value can be determined iteratively by varying transaction paths among Non TxNodes if there is no barrier in the path.

Based on the direct trust value (nodes 6 through 4) and recommended value, the DCT nodes (nodes 1 through 9) establish the local trust. The overall trust value is needed to increase a node's level of trust completely. Initially, each node's trust value was equal to 1/N, where N is the total number of nodes in the network system. A global trust value is required whenever a new block is added to the blockchain network. Algorithm 10 shows how the global trust value is calculated.

**Algorithm 10**. Calculation of Global Trust for Patient Enrollment Channel

Input: $node_i$, node set N*odes*
Output: Global trust value of $node_i$ (node-6)
1. $T_i \leftarrow 0$;
2. **For** $node_j$(1,2,3,4,5,6,7,8,9) ∈ *Nodes* **Do**
3.    $T_i = \sum C_{ji} T_j$;
4. **End**

According to Algorithm 10, $node_i$'s global trust value is calculated from the nodes already in existence and is the product of its local trust value and the corresponding global trust value of the other nodes to ensure the immutability of DCT transacting data on z-TAB.

*7.4. Transparency and accountability*

The clinical trial process involves transferring data from patients' wearable devices, categorized by zones (as per Table 9), to their respective PIs via the Merkel root. In the z-TAB model, American clinical study sites are assigned, where Principal Investigators Dr. Henri Johnson (patients AHJ1001, AHJ1002, ..., AHJ...n), Dr. Robert Kol (patients ARK1001, ARK1002, ..., ARK...n), and Dr. Smith (patients AS1001, AS1002, ..., AS...n) recruit patients. Wearable devices (WDs) are interconnected through the IoTs to the study teams, who verify the clinical data recorded in accordance with the approved study protocol.

A patient enrollment channel (comprising Node-6, Node-5, Node-4, and Node-2) remains active on the Hyperledger, with every Clinical Research Organization (CRO), PI, Ethics Committee (EC), and patient involved in the data flow, represented as hash values against recorded data in the z-TAB model. These active channel nodes monitor each patient's activity to authenticate data compliance with the approved protocol, ALCOA (Attributable, Legible, Contemporaneous, Original, and Accurate), ICH-GCP (International Conference on Harmonization - Good Clinical Practice), EC requirements, and applicable country-specific regulations. Transparency and data accountability are inherent within the Merkel root of the channelled active nodes, where these nodes are responsible for the recorded wearable device data and uphold transparency among all nodes within the patient enrollment channel (as delineated in Table 10).

**Table 10**
Data flow direction on the patient enrollment channel.

| Clinical Study Site No. | Patient (Node-6) | Wearable Devices Zones | PI (Node-5) | EC (Node-4) | Hash values | CRO (Node-2) |
|---|---|---|---|---|---|---|
| 1 | AHJ1001<br>AHJ1002<br>AHJ....n | "A" to "G"<br>"A" to "G"<br>"A" to "G" | Dr Henri Johnson | EC-1 | 9edbc9e799247cc5d54c00a7d999f7251641fe867cef4aaeebfff5aa84161e4a | CRO receives and verifies clinical data on the Patient enrollment channel through Hyperledger Fabric on the proposed z-TAB model |
| 2 | ARK1001<br>AHR1002<br>AHR....n | "A" to "G"<br>"A" to "G"<br>"A" to "G" | Dr. Robert Kole | EC-2 | 0339e39f671e14ef6a4215a93b44b4814bde387db579a67d3ac90eedf63a0c17 | |
| 3 | AS1001<br>AS1002<br>AS....n | "A" to "G"<br>"A" to "G"<br>"A" to "G" | Dr. Smith | EC-3 | 1630b66a773dc28bbec575a77036460c07e9796bffe6477b3454b25c7d706f66 | |

## 7.5. Tracking and tracing

The z-TAB model seamlessly enables comprehensive tracking and tracing of data flows across the Hyperledger Fabric system, effectively managing DCTs worldwide. Each data transaction occurs through patients' wearable devices, progressing to PIs, clinical research organizations (CROs), and other nodes in a timestamped manner on the blockchain platform. Recorded data at specific times can be accessed by active nodes within the patient enrollment channel (Node-6, Node-5, Node-4, and Node-2), where transactional information is visible to all nodes with synchronized updates.

Clinical data can be traced at any given moment to authenticate recorded and reported data within the respective active channel. Fig. 25 illustrates the data transaction process, Merkel root, and timestamping, facilitating the tracking and tracing of DCT shipments over time.

| Patient (Node-6) | | | PI (Node-5) | | | EC (Node-4) | | | CRO (Node-2) | | |
|---|---|---|---|---|---|---|---|---|---|---|---|
| Block Header | Previous Hash | Nonce | Block Header | e1fa31ce0f2cad03486aff30 31f178da6c2c3c57ed7e147 70a5742f69e4004e5 | Nonce | Block Header | 4f8d31ad5d44cc85a9ec814 9df1a137ff474c8650a7d41 011bc10495ad528b78 | Nonce | Block Header | 18d401268aff9ae17 c49b7e4a15994140 c61313818d809f8c | Nonce |
| | Time Stamp | Merkle Root | | Time Stamp | Merkle Root | | Time Stamp | Merkle Root | | Time Stamp | Merkle Root |
| Body | Data Transactions | | Body | Data Transactions | | Body | Data Transactions | | Body | Data Transactions | |
| | WD (A-G) | TX1…..TXn | | PI activities | TX1…..TXn | | EC activities | TX1…..TXn | | CRO activites | TX1…..TXn |
| SHA256-Hash | e1fa31ce0f2cad03486aff3031f178d a6c2c3c57ed7e14770a5742f69e400 4e5 | | SHA256-Hash | 4f8d31ad5d44cc85a9ec8149df1a137ff4 74c8650a7d41011bc10495ad528b78 | | SHA256-Hash | 18d401268aff9ae17c49b7e4a1599414 0c61313818d809f8cc2ce30f35e42891 | | SHA256-Hash | 226c5345f3c15842ec3ff34a7b92f126 1557170850d2cb0f3dde80030aade6e 6 | |

**Fig. 25.** Time stamp of patient enrollment channel nodes for tracking and tracing.

## 7.6. Temperature–humidity control

Throughout DCTs, investigational medicinal products (IMPs) and other biological specimens are transferred among nodes within the blockchain consortium. The Sponsor, typically a pharmaceutical company, dispatches IMPs to patients at various locations with randomized allocation. Simultaneously, the Sponsor arranges patient blood sample collection according to the protocol schedule. These samples must reach designated pathological laboratories without damage, spillage, loss, or deterioration, maintaining a set temperature and humidity-controlled conditions. In the z-TAB model, transactional information is continuously updated at each transfer point. Data loggers are affixed to IMP and blood sample transportation packages to monitor temperature and humidity control parameters under specific protocol conditions. Any alterations or deviations in these parameters during shipment are readily observable by active channel nodes within the Sponsor channel (Table 11).

**Table 11**
Temperature and humidity control on the sponsor channel

| Name of Channel | Nodes of the private channel | Active and inactive nodes on a private channel | Functions of Channel |
|---|---|---|---|
| Sponsor channel | Node-1,2,3,5,6 | Active: 1,2,3,5,6 Inactive:4,7,8,9 | Allocation of sponsor duties, Overall monitoring of DCTs, Country specific dispatch of IMPs, and Blood sample collection laboratory agreements |

| | Sponsor (Node-1) | | | CRO (Node-2) | | | Regulatory (Node-3) | |
|---|---|---|---|---|---|---|---|---|
| Block Header | Previous Hash | Nonce | Block Header | 8f2997c954e3cf1f3499ad4b728c93bf5b3ed0a0e6eb8a57f7e29f9bf8ee836f | Nonce | Block Header | d65e1e8a06eea8fce8a789888afd17ff900465cd2e1ada90221c11ceb928b3d2 | Nonce |
| | Time Stamp | Merkle Root | | Time Stamp | Merkle Root | | Time Stamp | Merkle Root |
| Body | Data Transactions | | Body | Data Transactions | | Body | Data Transactions | |
| | Temperature (°C) | TX1…..TXn | | Temperature (°C) | TX1…..TXn | | Temperature (°C) | TX1…..TXn |
| | Relative Humaidity (%RH) | TX1…..TXn | | Relative Humaidity (%RH) | TX1…..TXn | | Relative Humaidity (%RH) | TX1…..TXn |
| SHA256-Hash | 8f2997c954e3cf1f3499ad4b728c93bf5b3ed0a0e6eb8a57f7e29f9bf8ee836f | | SHA256-Hash | d65e1e8a06eea8fce8a789888afd17ff900465cd2e1ada90221c11ceb928b3d2 | | SHA256-Hash | 3a6fdd4a90f39bbd5ad5edb43879766bfe16ed58d5dfd0860f3b088f1fdcd295 | |

| | Patient (Node-6) | | | PI (Node-5) | |
|---|---|---|---|---|---|
| Block Header | 3a6fdd4a90f39bbd5ad5edb43879766bfe16ed58d5dfd0860f3b088f1fdcd295 | Nonce | Block Header | ef68cd175d797ed2e89bc85e936ca28c804211eb8acdb2b1709c2f27b65304cf | Nonce |
| | Time Stamp | Merkle Root | | Time Stamp | Merkle Root |
| Body | Data Transactions | | Body | Data Transactions | |
| | Temperature (°C) | TX1…..TXn | | Temperature (°C) | TX1…..TXn |
| | Relative Humaidity (%RH) | TX1…..TXn | | Relative Humaidity (%RH) | TX1…..TXn |
| SHA256-Hash | ef68cd175d797ed2e89bc85e936ca28c804211eb8acdb2b1709c2f27b65304cf | | SHA256-Hash | 2c51cac056a3e80df0576779d095b217ea513d6ec1785adff7c877320a542a85 | |

**Fig. 26.** Temperature and humidity control data records.

In Fig. 26, each node within the Sponsor channel (Node-1, Node-2, Node-3, Node-5, Node-6) possesses a ledger copy containing shipment data pertinent to DCT logistics. Maintaining controlled temperature and humidity levels during these shipments is paramount and rigorously monitored. Should deviations occur, the responsible party can be identified and recorded due to the immutable nature of blockchain technology, which prevents retrospective alterations. Every node receives transit updates, enabling prompt detection of any deviations that may impact the quality of shipments. Such deviations prompt immediate corrective and preventive actions by DCT stakeholders, particularly sponsors and Clinical Research Organizations (CROs), to ensure safer shipments. Thus, variations in climatic conditions (temperature/humidity) during shipments can be observed, monitored, and controlled by protocol requirements, enhancing the compliance and success of DCTs.

## 8. Conclusions, implications, and further recommendations

This research explores a z-TAB model, which integrates blockchain, Hyperledger Fabric, zero-trust principles, the IoT, and T-PBFT to facilitate DCTs worldwide. The primary objective of the z-TAB model is to streamline DCT data collection from thousands of patients automatically, eliminating the need for intermediaries. Data collection is seamlessly conducted through smart contracts, Hyperledger Fabric, zero-trust architecture, blockchain, and T-PBFT, enhancing data scalability on a global scale. T-PBFT is implemented in the z-TAB model to streamline the consensus process, reduce the number of consensus nodes and increase efficiency while mitigating communication complexities among nodes (Node-1 to Node-9), even if some nodes fail to achieve mutual consensus. Various policies (PESC, PIP, PDPSC, PEF, TESC, and ABAC) within the z-TAB model approve each DCT activity in which patients participate through wearable devices. The model's zero-trust architecture ensures that all data access is authenticated by private channel nodes, preventing intrusions or unauthorized access. To support the management of DCTs across nations, the model is evaluated based on immutability, privacy and security, mutual consensus, transparency, accountability, tracking and tracing, and temperature–humidity control parameters, ensuring its validation and authentication. The model guarantees comprehensive data access, timestamping, clinical data quality, correctness, and readability by ALCOA criteria as per the US FDA standards. A recommendation for model advancement includes developing a software-based prototype and validating the DCT process in specific clinical research units.

**Declaration of competing interest**
The authors declare that no financial interests or individual relationships appear to influence the research work reported in the paper.

**Funding statement**
There are no funding sources to disclose.

**Author contributions**
Ashok Kumar Peepliwal: DCT Conceptualization, Blockchain, and Hyperledger fabric systems, Merkle Tree development, Application and Evaluation of z-TAB, C
Hari Mohan Pandey and Sudhinder Chauhan: Literature search, Wearable Devices and IoT application in DCTs
Anand A. Mahajan: Introduction and Rational Zero-trust Architecture Blockchain Development
Surya Prakash and Vinesh Kumar: Smart contract application in model, manuscript review, proofreading, referencing
Rahul Sharma: EightTrust practical byzantine fault tolerance for mutual consensus of DCT nodes

**Abbreviations**

The following abbreviations were used in this research manuscript.

| | |
|---|---|
| ABAC | : Attribute-Based Access Control |
| ALCOA | : Attributable, Legible, Contemporaneous, Original and Accurate |
| BFT | : Byzantine Fault Tolerance |
| BP | : Blood Pressure |
| CDMS | : Clinical Data Management System |
| CDMS | : Clinical Data Management System |
| CRO | : Clinical Research Organization |
| D2D | : Device to Device |
| DCT | : Decentralized Clinical Trial |
| DLT | : Distributed Ledger Technology |
| EC | : Ethics Committee |
| ECG | : Electro Cardio Gram |
| eCOA | : Electronic Clinical Outcome Assessment |
| EDC | : Electronic Data Capture |
| FTP | : File Transfer Protocol |
| Hash | : The act of creating a fixed-size output from a variable-sized input by applying the hash mathematical formulas is referred to as "hashing." |
| HTTPS | : Hypertext Transfer Protocol System |
| ICH-GCPs | : International Council for Harmonization-Good Clinical Practices |
| IMP | : Investigation medicinal Product |
| IoT | : Internet of Things |
| IPFS | : Interplanetary File System |
| Merkle Tree | : a hash tree with typically a branching factor of 2 (2 nodes) |
| MSP | : Membership Service Providers |
| Nonce | : a nonce is an arbitrary number used once in a cryptographic communication. |
| PBFT | : Practical Byzantine Fault Tolerance |
| PDP | : Policy Decision Point |

| | | |
|---|---|---|
| PDPSC | : | Policy Decision Point Smart Contract |
| PE | : | Policy Enforcement |
| PEP | : | Policy Enforcement Policy |
| PESC | : | Policy Enforcement Smart Contract |
| PI | : | Principal Investigator |
| PoET | : | Proof of Elapsed Time |
| POS | : | Proof of Stake |
| POW | : | Proof of Work |
| SHA-256 | : | Secure Hash Algorithm-256 |
| TESC | : | Trust Engine Smart Contract |
| Timestamp | : | a digital record of the time of occurrence of a particular event |
| T-PBFT | : | EigenTrust-Based Practical Byzantine Fault Tolerance |
| WD | : | Wearable Device |
| ZT | : | Zero-Trust |
| z-TAB | : | Name of proposed architecture model (Zero-trust architecture blockchain) |